\documentclass[a4paper,11pt]{article}

\pdfoutput=1

\usepackage{appendix}
\usepackage{jheppub,bm,slashed} 
\usepackage[legacycolonsymbols]{mathtools}
\usepackage[usenames,dvipsnames,svgnames,table,x11names]{xcolor}
\usepackage{extarrows}
\usepackage{easybmat}

\usepackage[normalem]{ulem}

\usepackage[T1]{fontenc}
\usepackage[utf8]{inputenc}

\newcommand{\tchar}[2]{%
  \vartheta\genfrac{[}{]}{0pt}{}{#1}{#2}%
}

\newcommand{\chvec}[2]{%
  \genfrac{[}{]}{0pt}{}{#1}{#2}%
}

\usepackage{array}
\usepackage{amsmath}

\usepackage{tikz}
\usepackage{pgfplots}
\pgfplotsset{compat=1.14}
\usepackage{tikz-3dplot}
\usetikzlibrary{intersections, calc,positioning,decorations.pathreplacing,decorations.pathmorphing,arrows,arrows.meta,patterns,pgfplots.fillbetween,math,matrix}

\usepackage{newtxtext}
\usepackage{bm}
\usepackage{amsfonts}
\usepackage{amssymb}
\usepackage{latexsym}
\usepackage{graphicx}
\usepackage{float}
\usepackage{siunitx}
\usepackage{braket}
\usepackage{physics}
\usepackage{cancel}
\usepackage{ascmac}
\usepackage{comment}

\usepackage{booktabs}
\usepackage{arydshln}
\usepackage[table]{xcolor}

\usepackage{longtable,booktabs,array}

\newcommand{\C}{\mathbb{C}}

\renewcommand{\H}{\mathcal{H}}

\newcommand{\R}{\mathbb{R}}

\newcommand{\Z}{\mathbb{Z}}

\renewcommand{\Im}{\mathrm{Im}}

\newcommand{\Spin}{\mathrm{Spin}}

\usepackage{cancel}

\newcommand{\q}{q^{L_0}\bar q^{\bar{L}_0}}
\newcommand{\ve}{\varepsilon}

\usepackage{booktabs}
\usepackage{tabularx}
\usepackage{array}
\usepackage{makecell}

\usepackage{nicematrix}

\usepackage[whole]{bxcjkjatype} 

\usepackage[T1]{fontenc} 
\usepackage[utf8]{inputenc}

\definecolor{light_blue}{rgb}{0.15, 0.35, 0.95}
\definecolor{kit_green}{rgb}{0, 
0.58823 
, 0.50980 
}

\usepackage{tikz}
\usepackage{pgfplots}
\pgfplotsset{compat=1.14}
\usepackage{tikz-3dplot}
\usetikzlibrary{intersections, calc,positioning,decorations.pathreplacing,decorations.pathmorphing,arrows,arrows.meta,patterns,pgfplots.fillbetween,math,matrix}

\usepackage{multirow}
\usepackage{booktabs} 
\usepackage{siunitx} 
\usepackage{tabularx} 
\usepackage{tikz-cd}

\renewcommand{\i}{\mathrm{i}}

\newcommand{\Ash}[1]{\mathsf A_{#1}}
\newcommand{\Bsh}[1]{\mathsf B_{#1}}
\newcommand{\Uone}{\mathrm{U}(1)}

\preprint{
\begin{minipage}{5cm}
\flushright
\end{minipage}} 

\title{Heterotic strings on Enriques surfaces}

\author[1]{Arata Ishige,}
\author[2]{Elisa Iris Marieni}

\affiliation[1]{Graduate Institute for Advanced Studies, SOKENDAI, 1-1 Oho, Tsukuba, Ibaraki 305-0801, Japan}
\affiliation[2]{STAG Research Center \& Mathematical Sciences, University of Southampton\\Highfield, Southampton SO17 1BJ, UK}

\emailAdd{arata@post.kek.jp}
\emailAdd{e.i.marieni@soton.ac.uk}

\abstract{We study orbifold compactifications of heterotic strings on
Enriques surfaces. We classify the inequivalent shift vectors for both
the \(E_8\times E_8\) and \(\mathrm{Spin}(32)/\mathbb Z_2\) lattices, and
analyse the light spectrum of the resulting models. We show that these
models can be interpreted as compactifications of ten-dimensional
non-supersymmetric heterotic strings on Enriques surfaces. For certain
classes of shifts, the moduli-independent tachyons inherited from the
parent theories are projected out.}

\begin{document}
\maketitle

\section{Introduction and Summary}

Non-supersymmetric aspects of superstring theory have been attracting growing attention in recent years \cite{Acharya:2022shu,DeFreitas:2024ztt,fraiman2025symmetriesdualitiesnonsupersymmetricchl,Kaidi_2021,GarciaEtxebarria:2020xsr,hamada20258dnonsupersymmetricbranesheterotic,Hamada:2024cdd,Acharya:2019mcu,Tachikawa:2024ucm,Tachikawa:2023lwf,Boyle_Smith_2024,Ishige:2026ihp,nakajima2023newnonsupersymmetricheteroticstring,Koga:2022qch,Baykara:2024tjr,Robbins:2025wlm,Leone:2025eht,Basile:2025mnj,Basile:2026lyc,Ooguri:2016pdq,ValeixoBento:2025qih,ValeixoBento:2025yhz,Larotonda:2026hxy}. This growing interest is motivated not only by the phenomenological observation that supersymmetry has yet to be discovered in nature, but also by the increasingly prominent role of non-supersymmetric systems in the swampland program on quantum gravity \cite{mcnamara2019cobordismclassesswampland,Kaidi:2023tqo,Dierigl_2023,Fukuda:2024pvu,Kaidi:2024cbx,Basile:2023knk}.

Indeed, the construction and classification of examples, the exploration of dualities among them, and even attempts to lift them to M-theory have a long history \cite{Dixon:1986iz,Sugimoto_1999,sagnotti1995propertiesopenstring,Kawai:1986vd,ALVAREZGAUME1986155,Klebanov_1999,Bergman:1999km,Fabinger_2000,Meessen_2001,Russo:2001tf}. However, the absence of supersymmetry, and hence of protected objects such as BPS states, often makes such attempts difficult.

Why have dualities in supersymmetric string theory worked so remarkably well? One reason is the existence of K3 surfaces and their rich geometry \cite{Witten_1995, Aspinwall:1994rg,Kiritsis_2000,aspinwall1999k3surfacesstringduality,Schwarz:1995bj,Duff:1996rs,Aldazabal:1996fm,witten1995commentsstringdynamics}. Several non-perturbative aspects of Type II strings in higher dimensions come down to K3 geometry: e.g. the realization of non-abelian gauge symmetries in terms of Kodaira singularities \cite{Vafa_1996}. K3 surfaces also shed light on string dualities in lower dimensions through K3-fibered Calabi-Yau threefolds \cite{Morrison:1996na,Morrison:1996pp,Kachru:1995wm}. 

There are non-supersymmetric analogues of K3 surfaces, called \textit{Enriques surfaces}, which can be obtained as fixed-point-free $\Z_2$ quotients of K3 surfaces. Since this is not Calabi–Yau, it breaks all supersymmetry. Enriques surfaces have often appeared in string-theoretic contexts before, but the constructions were often arranged so as to preserve supersymmetry \cite{Bershadsky:1998vn,Cheng:2023owv,Sharpe:2006qd,Sharpe:2013bwa,Eberhardt:2017uup,ArabiArdehali:2021iwe,Ferrara:1995yx,Klemm:2005pd}. 
Recently, Type IIA and IIB strings (more precisely Type 0A and Type 0B) on Enriques have been studied in \cite{Ishige:2026ihp}, where they have been conjectured to be dual to an asymmetric orbifold of heterotic strings, as non-supersymmetric extensions of known dualities in six and five dimensions. 

In the present work, we study heterotic strings on an Enriques surface with 48 inequivalent choices of shift vectors. We construct the worldsheet theory by taking an Enriques orbifold of supersymmetric heterotic strings on a K3 surface, itself realised at an orbifold point. We then show that the resulting models are closely related to ten-dimensional non-supersymmetric heterotic strings \cite{Dixon:1986iz,ALVAREZGAUME1986155}. We evaluate the light spectrum of these theories focusing on massless states and tachyons. 
Although the only tachyon-free parent 10-dimensional theory is the $D_8 \times D_8$, also some of the theories related to $E_8 \times D_8, (E_7 \times A_1)^2, A_{15}\times U(1)$ and $D_4\times D_{12}$ heterotic strings do not inherit the parent's tachyons. By contrast, all the compactifications descending from the $D_{16}$ theory retain tachyonic states.

This paper is organized as follows. In section \ref{sec: het on enriques} we examine heterotic strings on an Enriques surface. We describe the Enriques involution and use it to define an orbifold action. We construct the partition function and classify inequivalent shift vectors. In section \ref{sec: light spectrum} we analyse the light spectrum of the theory. In particular, we show that for a class of shift vectors, all moduli-independent tachyons are removed. Additional material is collected into appendices.

\section{Heterotic strings on Enriques surfaces}\label{sec: het on enriques}

We consider the 1-loop partition function of heterotic strings on Enriques surfaces. In our conventions the left-moving sector contains the superstring and the right-moving sector contains the bosonic string.

\subsection{Heterotic strings on \texorpdfstring{$T^4$}{T^4} (sixteen supercharges)}
In this section we review supersymmetric heterotic strings compactified on $T^4$. Conventions for the Jacobi theta functions can be found in Appendix \ref{sec:theta functions}.
The Hilbert space of heterotic strings on $T^4$ is
\begin{equation}
\begin{aligned}
\H_{T^4}=&\H_{\text{Boson}}^{4,4}\otimes\H_{\text{Fermion}}^{8,0}\otimes\H_{\Gamma_{4,20}},\\
    \H_{\Gamma_{4,20}}=&\H_{\text{Boson}}^{4,20}\otimes \qty(\bigoplus_{p\in\Gamma_{4,20}}\C\ket{p}).
\end{aligned}
\end{equation}
To compute the partition function, let us work at a special point in moduli space, where we can take
\begin{align}
    \Gamma_{4,20} = \Gamma_{4,4} \oplus \Gamma_{0,16} \, ,
\end{align}
so that the two lattices are separately self-dual and have no mixing. Here, $\Gamma_{4,4}=H_1(T^4)\oplus H^1(T^4)$ is a $(4,4)$ even self-dual lattice. We will use the same symbol for the lattice and its theta series when no confusion can arise.
The partition function is
\begin{equation}\label{eq:hetonT4}
\begin{aligned}
      &Z_{T^4}=\frac{1}{(\Im \tau)^2\eta^4\bar\eta^4}\frac{\Theta}{\eta^4} \frac{\Gamma_{4,4}}{\eta^4\bar\eta^{4}}\frac{\bar\Gamma_{16}}{\bar\eta^{16}},\\
         \Theta(q)=&\qty(\sum_{r\in V}-\sum_{r\in Sp})q^{\frac{1}{2}r^2},\quad
      \Gamma_{4,4}=\sum_{p \in \Gamma_{4,4}}q^{\frac12p_L^2} \bar q^{\frac12p^2_R},\quad
    \bar\Gamma_{16}=\sum_{p\in\Gamma_{0,16}}\bar q^{\frac{1}{2}p^2}
\end{aligned}
\end{equation}
where $\Theta$ comes from left-moving fermions, and $V,Sp$ are the vector and spinor lattices of $SO(8)$:
\begin{equation}\label{eq:V_and_Sp}
    \begin{aligned}
        V=&\left\{(r_1,\cdots,r_4)\in\Z^4\middle|\sum_{i=1}^4 r_i\in2\Z+1\right\},\\
        Sp=&\left\{(r_1,\cdots,r_4)\in\qty(\Z+\frac12)^4\middle|\sum_{i=1}^4 r_i\in2\Z +1\right\}.\\
    \end{aligned}
\end{equation}
 The last factor $\bar\Gamma_{16}$ comes from internal right-moving bosons and should be even and self-dual for modular invariance. Then it is restricted to one of the even self-dual lattices, $E_8\times E_8$ or $\Spin(32)/\Z_2$:
\begin{equation}\label{eq: lattices}
\begin{aligned}
     E_8=&\left\{p\in\Z^8 \text{ or }  (\Z+\tfrac{1}{2})^8\middle| \sum_{i=1}^8 p_i\in2\Z\right\} \, ,\\
     \Spin(32)/\Z_2=&\left\{p\in\Z^{16}\text{ or }\qty(\Z+\tfrac{1}{2})^{16}\middle|\sum_{i=1}^{16}p_i \in2\Z\right\} \, ,
\end{aligned}
\end{equation}
Modular transformations of each factor are given by:
\begin{equation}
    \begin{aligned}
    T\cdot \frac{1}{\eta^{12}\bar\eta^{24}}=&(-1)\frac{1}{\eta^{12}\bar\eta^{24}},&&S\cdot \frac{1}{(\Im \tau)^2\eta^4\bar\eta^4}=\frac{1}{(\Im \tau)^2\eta^4\bar\eta^4}\\
        T\cdot \Theta=&(-1)\Theta, &&S\cdot \frac{\Theta}{\eta^4}=\frac{\Theta}{\eta^4},\\
        T\cdot\Gamma_{4,4}=&\Gamma_{4,4},&&S\cdot \frac{\Gamma_{4,4}}{\eta^4\bar\eta^4}=\frac{\Gamma_{4,4}}{\eta^4\bar\eta^4}\\
        T\cdot  \bar\Gamma_{16}=& \bar\Gamma_{16} , 
         &&S\cdot \frac{\bar\Gamma_{16}  }{\bar\eta^{16}}=\frac{\bar\Gamma_{16} }{\bar\eta^{16}},
    \end{aligned}
\end{equation}
Therefore, $Z_{T^4}$ is modular invariant.

\subsection{Enriques involution}
Starting from heterotic strings on $T^4$, a singular Enriques surface can be obtained as the orbifold limit of subsequent $\mathbb{Z}_4$ and $\mathbb{Z}_2$ actions. This can be thought of as the "orbifold of an orbifold" \cite{DIXON1985678,DIXON1986285,Narain:1986qm}. Let $g$ denote the orbifold generator of the $\mathbb{Z}_2$ action and $h$ the orbifold generator of the $\mathbb{Z}_4$ action, then
\begin{align}
    g: \begin{cases}
        X^6 \rightarrow-X^6\\
        X^7 \rightarrow -X^7\\
        X^8 \rightarrow-X^8\\
        X^9 \rightarrow-X^9
    \end{cases} \, , \quad \begin{cases}
        \psi^6 \rightarrow-\psi^6\\
        \psi^7 \rightarrow -\psi^7\\
        \psi^8 \rightarrow -\psi^8\\
        \psi^9 \rightarrow -\psi^9
    \end{cases} \, ;&\quad \quad h: \begin{cases}
        X^6 \rightarrow-X^6\\
        X^7 \rightarrow -X^7+\pi R_7\\
        X^8 \rightarrow X^8\\
        X^9 \rightarrow X^9 + \pi R_9,
    \end{cases} \, , \quad \begin{cases}
        \psi^6 \rightarrow -\psi^6\\
        \psi^7 \rightarrow -\psi^7\\
        \psi^8 \rightarrow \psi^8\\
        \psi^9 \rightarrow \psi^9
    \end{cases} \, , \\
    &[g,h]=0 \, , \nonumber
\end{align}
where $X^i+2\pi R_i=X^i$ with $i=6,...,9$ are bosonic compact coordinates, $\psi^{i}$ are worldsheet fermions and all other coordinates are left untouched. It can be easily seen that the action of $g$ on $T^4$ provides $16$ fixed points, and the quotient $T^4/g$ can be viewed as an orbifold limit of a $K3$ surface. It is also easily seen that $h$ acts freely on $T^4/g$, and it exchanges singular points pairwise. It should be noted that although $h^2$ is the identity on K3 surfaces, it lifts to $-1$ on the spinor bundle of $K3$. 

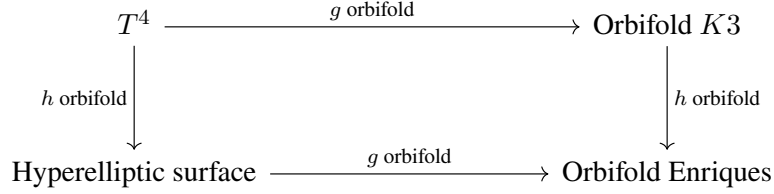
\begin{figure}[h]
    \centering
    \begin{tikzcd}[row sep=huge, column sep=huge]
T^4 
  \arrow[rr, rightarrow, "g \text{ orbifold}" ] 
  \arrow[d, "h \text{ orbifold}"']
&&
\text{Orbifold }K3 
  \arrow[d, "h\text{ orbifold}"]
\\
\text{Hyperelliptic surface}
  \arrow[rr, rightarrow, "g  \text{ orbifold}"] 
&&
\text{Orbifold Enriques} 
\end{tikzcd}\caption{Four-dimensional manifolds}\label{fig:relevant_manifolds}
\end{figure}

The relation between these manifolds is schematically drawn in Figure \ref{fig:relevant_manifolds}.
For later use, it will be useful to define the same actions on bosonised fermions. Let $r$ be a spinor or vector weight of $SO(8)$, then
\begin{align}
    g\ket{r} = e^{2 \pi \i r \cdot v_{1f}} \ket{r} \, , \quad h\ket{r}= e^{2 \pi \i r\cdot v_{2f}} \ket{r} \, \\
    v_{1f}= \frac{1}{2}(1,1,0,0) \, , \quad v_{2f} = \frac12 (1,0,0,0)  \, .
\end{align}
Here $v_{1f}$ flips four fermions $\psi^6,\cdots,\psi^9$, while $h$ flips just two fermions $\psi^6,\psi^7$. Our orbifold actions leave the rank 16 lattice completely unrotated, so that the invariant sublattices $I\subset \Gamma_{4,4}$ with respect to each action have signatures
\begin{align}
    \text{sig}\left (I_g \right) = (0,0) \, , \quad \text{sig}\left(I_{h}\right) = (2,2) \, ,
\end{align}
where $\text{vol}(I_h)=2$\footnote{The $\mathbb{Z}_4$ orbifold is part of a class of orbifolds constructed in \cite{Acharya:2022shu} with an additional spectator $S^1$. In their notation, this corresponds to the $(r,a,\delta)=(18,2,0)$ point in the Nikulin involution space. These orbifolds act as $\mathbb{Z}_2$ on bosons and as $\mathbb{Z}_4$ on fermions.}. The actions of $g,h$ on $\mathcal{H}_{\Gamma_{4,4}}$ are explained in Appendix \ref{sec:Twisted bosonic fock space}.

The orbifold action on $\Gamma_{0,16}$ is defined up to a phase depending on a shift vector. Since the gauge lattice is left geometrically untouched, these shifts can be chosen, without loss of generality, to lie entirely along the $\Gamma_{0,16}$ directions. We will denote by $v,w$ the shift vectors related to the $\mathbb{Z}_2,\mathbb{Z}_4$ action respectively:
\begin{equation}
    \begin{aligned}
        g\ket{p}=&e^{-2\pi ip\cdot v}\ket{p}\,,\\
        h\ket{p}=&e^{-2\pi ip\cdot w}\ket{p}\, ,
    \end{aligned}
\end{equation}
for $p\in\Gamma_{0,16}$.
\subsection{Construction}

The partition function of heterotic strings on an Enriques surface is constructed as 
\begin{equation}
\begin{aligned}
    Z_{\text{Enriques}}
     =&\frac{1}{(\Im \tau)^2\eta^4\bar\eta^4}\frac{1}{\eta^8 \bar \eta^{20}}\frac{1}{4}\sum_{k,l=0}^3\frac{1}{2}\sum_{c,d=0}^1Z^{k,l}\chvec{c}{d}\\
     Z^{k,l} \chvec{c}{d} =& \tr_{\mathcal{H}^{g^ch^k}} g^d h^l q^{L_0} \bar q^{\bar L_0}\\
     =&\Theta^{k,l}\chvec{c}{d}\Gamma^{k,l}_{4,4}\chvec{c}{d} \overline{\Gamma}_{16}^{k,l}\chvec{c}{d} ,  \label{eq: op}
\end{aligned}
\end{equation}
where we have separated contributions from fermions, rank 8 and rank 16 lattices. Here $\mathcal{H}^{g^ch^k}$ denotes the Hilbert space of the $g^ch^k$-twisted sector, which means that boundary conditions of fields along spatial $S^1$ on $T^4$ are twisted by $g^ch^k$. 
The explicit expressions of these building blocks are\footnote{We can include an additional phase $e^{-\pi i(dk-lc)N}$ for $N=0,1$, in the partition function; this is the discrete torsion phase \cite{Vafa:1986wx}. In the present case, however, it does not lead to a different spectrum, because the only modular orbit on which this phase is non-trivial has a vanishing $\Gamma_{4,4}^{k,l}\chvec{c}{d}$
 contribution. In the present paper we focus on $N=0$.}
\begin{align}\label{eq: enriques pf pieces}
    \Theta^{k,l}\chvec{c}{d} 
    =&e^{-\pi \i (dv_{1f} +lv_{2f})\cdot (cv_{1f}+kv_{2f})} \left(\sum_{r \in V} - \sum_{r \in Sp} \right)q^{\frac12(r+cv_{1f}+kv_{2f})^2}e^{2 \pi \i (r+cv_{1f}+kv_{2f}) \cdot (d v_{1f}+ lv_{2f})} \, , \nonumber\\
    \overline{\Gamma}^{k,l}_{16}\chvec{c}{d} =& e^{\pi \i (dv+lw)\cdot (cv+kw)} \sum_{p \in \Gamma_{0,16}} \bar q^{\frac12 (p+cv+kw)^2}e^{-2 \pi \i (p+cv + kw)\cdot (dv + lw)} \, ,
\end{align}
where the form of $\Gamma_{4,4}^{k,l}\chvec{c}{d}$ is discussed in Appendix \ref{sec:Twisted bosonic fock space}. The shift vectors are subject to the following constraints
\begin{align}\label{eq: shift vec cond}
    2 v \in \Gamma_{0,16} \, , \quad 4w \in \Gamma_{0,16} \, , \quad 2v^2 \in 2 \mathbb{Z}+1 \, , \quad 4w^2 \in 2 \mathbb{Z}+1\, , \quad v \cdot 4w \in 4\mathbb{Z}+1 \, .
\end{align}
These conditions follow from imposing invariance of $Z^{k,l}\chvec{c}{d}$ under $d\to d+2$ and $l\to l+4$ , which comes from $g^2=1$ and $h^4=1$:
\begin{align}
e^{-2\pi \i(2v\cdot p)}e^{-\pi \i c(2v^2+1)} e^{-\pi \i k \left(2v\cdot w -\frac{1}{2} \right)}=&1,\\
    e^{-2\pi \i(4w\cdot p)}e^{-\pi \i c(1+4v\cdot w)}e^{-\pi \i k (1+4w^2)}=&1,
\end{align}
for every $c,k\in\Z$ and $p\in\Gamma_{0,16}$. Here we used the invariance of $\overline \Gamma_{4,4}^{k,l}\chvec{c}{d}$ under $d\to d+2$ and $l\to l+4$, which can be explicitly checked by using the expressions in Appendix \ref{sec:Twisted bosonic fock space}.

A comment on the choice of phases in \eqref{eq: enriques pf pieces} is in order. In the purely $g$ or $h$ orbifolds, when $k,l=0$ or $c,d=0$, the reader will recognise the usual orbifold actions described in the previous section. However, in "mixed" terms, new phases appear. For instance, in the $g$-twisted sector the action of $h$ is defined up to an additional phase $e^{\pi \i v \cdot w}$:
\begin{align}
\overline{\Gamma}^{0,1}_{16}\chvec{1}{0}=e^{\pi \i v \cdot w} \sum_{p \in \Gamma_{0,16}} \bar q^{\frac12 (p+v)^2} e^{-2 \pi \i (p+v) \cdot w} \, .
\end{align}
The phases are chosen so that each piece of the partition function has the following transformation properties under the modular group
\begin{equation}\label{eq: STtransf}
    \begin{aligned}
       T\cdot \Theta^{k,l}\chvec{c}{d}=&(-1)\Theta^{k,k+l}\chvec{c}{c+d},&&S\cdot \frac{\Theta^{k,l}\chvec{c}{d}}{\eta^4}=\frac{\Theta^{l,-k}\chvec{d}{-c}}{\eta^4},\\
        T\cdot\Gamma^{k,l}_{4,4}\chvec{c}{d}=&\Gamma^{k,k+l}_{4,4}\chvec{c}{c+d}, &&S\cdot\frac{\Gamma^{k,l}_{4,4}\chvec{c}{d}}{\eta^4\bar\eta^4}=\frac{\Gamma^{l,-k}_{4,4}\chvec{d}{-c}}{\eta^4\bar\eta^4},\\
T\cdot  \overline{\Gamma}^{k,l}_{0,16}\chvec{c}{d}=& \overline{\Gamma}^{k,k+l}_{0,16}\chvec{c}{c+d} ,
         &&S\cdot \frac{1}{\bar\eta^{16}}\overline{\Gamma}^{k,l}_{0,16}\chvec{c}{d}  =\frac{1}{\bar\eta^{16}}\overline{\Gamma}^{l,-k}_{0,16}\chvec{d}{-c} \, .
    \end{aligned}
\end{equation}
Then
\begin{equation}
\begin{aligned}
        T\cdot Z^{k,l}\chvec{c}{d}=&Z^{k,k+l}\chvec{c}{c+d},\\
        S\cdot  Z^{k,l}\chvec{c}{d}=&Z^{l,-k}\chvec{d}{-c},
\end{aligned}
\end{equation}
and the total partition function of heterotic strings on Enriques is modular invariant.

The $4^22^2=64$ pieces of the partition function can be organised into 8 orbits with respect to $T,S$ transformations. Denoting each orbit by its generator:
\begin{enumerate}
    \item $Z^{0,0}\chvec{0}{0}$: 1 term; this is the partition function of heterotic strings on $T^4$ \eqref{eq:hetonT4};
    \item $Z^{0,0}\chvec{0}{1}$: 3 terms; these are part of the partition function of heterotic strings on K3 $=T^4/g$;
    \item $Z^{0,2}\chvec{0}{0}$: 3 terms; these terms correspond to a  direct compactification of 10d non-supersymmetric heterotic strings \cite{Dixon:1986iz} on $T^4$;
    \item $Z^{0,1}\chvec{0}{0}$: 12 terms; these terms together with the ones in orbit 3 are part of the partition function of heterotic strings on hyperelliptic surface $T^4/h$ \cite{Acharya:2022shu};
    \item $Z^{0,1}\chvec{0}{1}$: 12 terms.
    \item $Z^{0,2}\chvec{0}{1}$: 3 terms.
     \item $Z^{0,2}\chvec{1}{0}$: 6 terms.
    \item $Z^{0,1}\chvec{1}{0}$: 24 terms; all of these vanish because $\Gamma_{4,4}^{k,l}\chvec{c}{d}=0$ in this orbit; see Appendix \ref{sec:Twisted bosonic fock space}.
\end{enumerate}

\subsection{Enriques shift vectors}\label{sec: enriques shift vectors}
Using the conditions derived in \eqref{eq: shift vec cond}, it is possible to classify inequivalent shift vectors.
The complete derivation is presented in Appendix \ref{sec:classification}, here we summarise the final result. Throughout this section we fix the $\Z_2$ shift to be
\begin{align}
    v=\frac12(0^{14},1,1) \, ,
\end{align}
in the canonical basis.

\subsubsection{\texorpdfstring{$E_8\times E_8$}{E8×E8} shift classes}
Classification of shift vectors of $E_8\times E_8$ is summarized in Appendix \ref{sec:classification}. As a result, we have $24$ inequivalent shifts:

\begin{equation}
\begin{aligned}
\text{$E_8\times D_8$}:
&\quad \{\Ash{1},\Ash{2},\Ash{3}\}\times\{\Bsh{1},\Bsh{2}\}
,\quad
\{\Ash{4},\Ash{5}\}\times\{\Bsh{3},\Bsh{4}\},\\
\text{$D_8\times D_8$}:
&\quad \{\Ash{4},\Ash{5}\}\times\{\Bsh{5}\}
,\quad
\{\Ash{6}\}\times\{\Bsh{1},\Bsh{2}\},\\
\text{$(E_7\times A_1)^2$}:
&\quad \{\Ash{7},\Ash{8}\}\times\{\Bsh{6},\Bsh{7},\Bsh{8}\}
,\quad
\{\Ash{9},\Ash{10}\}\times\{\Bsh{9},\Bsh{10}\}.
\end{aligned}
\end{equation}
Here $E_8\times D_8, D_8\times D_8$ and $(E_7\times A_1)^2$ denote the corresponding ten-dimensional non-supersymmetric parent heterotic theories.

Here \(\mathsf A_i\) and \(\mathsf B_j\) denote the components of the shift
\(w=(\mathsf A_i: \mathsf B_j)\in \frac14(E_8\times E_8)\) in the first and second
\(E_8\) factors, respectively. Their unbroken algebras are summarized as

\renewcommand{\arraystretch}{1.15}
\begin{equation}
\begin{array}{@{}l@{\qquad}l@{\qquad\qquad}l@{\qquad}l@{}}
\mathsf A_1: E_8
&
\mathsf B_1: D_4\times A_1^2\times U(1)^2
&
\mathsf A_6: A_7\times U(1)
&
\mathsf B_6: A_5\times U(1)^3
\\
\mathsf A_2: E_7\times A_1
&
\mathsf B_2: D_6\times U(1)^2
&
\mathsf A_7: E_7\times U(1)
&
\mathsf B_7: D_6\times A_1\times U(1)
\\
\mathsf A_3: D_8
&
\mathsf B_3: E_6\times U(1)^2
&
\mathsf A_8: D_6\times A_1\times U(1)
&
\mathsf B_8: E_7\times U(1)
\\
\mathsf A_4: D_7\times U(1)
&
\mathsf B_4: A_7\times U(1)
&
\mathsf A_9: E_6\times A_1\times U(1)
&
\mathsf B_9: D_5\times A_1\times U(1)^2
\\
\mathsf A_5: D_5\times A_3
&
\mathsf B_5: A_5\times A_1\times U(1)^2
&
\mathsf A_{10}: A_7\times A_1
&
\mathsf B_{10}: A_3^2\times A_1\times U(1).
\end{array}
\label{eq:list_of_AB}
\end{equation}

For given $v,w$ and some original root system $R$ the root system of the unbroken gauge algebra $R_{v,w}$ is 
\begin{equation}
    R_{v,w}=\{\alpha\in R|\alpha\cdot v,\alpha\cdot w\in\Z\} \, .
\end{equation}
Since roots transform under the orbifold action with phases $e^{2 \pi \i \alpha \cdot v},e^{2 \pi \i \alpha \cdot w}$, $R_{v,w}$ is the sub-system of roots left invariant by the shifts $v,w$.

\subsubsection{\texorpdfstring{$\Spin(32)/\Z_2$}{Spin(32)/Z2} shift classes}
Similarly, there are $24$ inequivalent $w$-shifts on $\Spin(32)/\Z_2$. 
In the case of $4w\in \Z^{16}$, there are $10+8+2=20$ shift vectors:
\begin{equation}
    \begin{aligned}
    W_{0,\frac12}(n_1,n_2)=&
    \qty(\frac{1}{4}\qty(0^{14-n_1-n_2},1^{n_1},2^{n_2}),0,\frac{1}{2}):\\
    (n_1,n_2)\in\Bigl\{
(0,0),(0,2)&,(0,4),(0,6),
(4,1),(4,3),(4,5),
(8,0),(8,2),(12,1)
\Bigr\},\\
    \end{aligned}
\end{equation}
and
\begin{equation}
    \begin{aligned}
        W_{\frac14,\frac14}(n_1,n_2)=&\qty(\frac{1}{4}\qty(0^{14-n_1-n_2},1^{n_1},2^{n_2}),\frac14,\frac14):\\
(n_1,n_2)\in\Bigl\{
(2,0),(2,2)&,(2,4),(2,6),
(6,1),(6,3),(10,0),(10,2)
\Bigr\} \, ,
    \end{aligned}
\end{equation}
and
\begin{equation}
    \widetilde{W}_{\frac14,\frac14}(n)=\qty(-\frac12,\frac12^{11-8n},\frac14^{2+8n},\frac14,\frac14),\quad n=0,1 \, .
\end{equation}
The unbroken gauge algebras can be identified by using Weyl subgroup of $D_{14}\times A_1$ which stabilises $W_{\ast,\ast}(n_1,n_2)$, as follows:
\begin{equation}
\begin{aligned}
    W_{0,\frac12}(n_1,n_2)&:\quad
D_{14-n_1-n_2}\times A_{n_1-1}\times D_{n_2}\times \Uone^{3},\\
W_{\frac14,\frac14}(n_1,n_2)&:\quad
D_{14-n_1-n_2}\times A_{n_1-1}\times A_1\times D_{n_2}\times U(1)^{2},\\
\widetilde{W}_{\frac14,\frac14}(n)&:D_{12-8n}\times A_{1+8n}\times A_1\times U(1)^2
\end{aligned}
\end{equation}
Here, $D_1$ means $U(1)$ and $A_{-1}\times U(1)^3$ means $U(1)^2$.

The shifts in the odd class are
\begin{equation}
\begin{aligned}\label{eq:new_shifts}
W_{\frac18,\frac38}(n)=&\left(\qty(\frac18)^{11-4n},\qty(\frac38)^{3+4n},\frac18,\frac38\right) ,\quad n=0,1,\\
    W_{\frac58,-\frac18}(n)=&
\left(
\left(\frac18\right)^{13-4n},
\left(\frac38\right)^{1+4n},
\frac58,
-\frac18
\right),
\quad n=0,1,
\end{aligned}
\end{equation}
with gauge algebras
\begin{equation}
\begin{aligned}
        A_{10-4n}\times A_{2+4n}\times U(1)^{4},\\
        A_{12-4n}\times A_{4n}\times U(1)^4,
\end{aligned}
\end{equation}
respectively.
These shift vectors for $\Spin(32)/\Z_2$ with their parent non-supersymmetric heterotic strings and gauge algebra are summarised in table \ref{tab: massless so32}.

\subsubsection{Parent non-susy heterotic strings}

Heterotic string theories without spacetime supersymmetry in 10 dimensions can be constructed as $\Z_2$ asymmetric orbifolds of supersymmetric ones \cite{Dixon:1986iz}. These orbifolds act only on the internal compact degrees of freedom and hence do not reduce the number of spacetime dimensions. In total, there are seven inequivalent theories corresponding to seven inequivalent shift vectors $\delta \in \frac12 \Gamma_{0,16}$. The shift vectors fall into two classes: $\delta^2=1$ giving rise to tachyonic models and $\delta^2=2$ that lead to tachyon-free theories. For $\Spin(32)/\Z_2$, the inequivalent shifts are
\begin{equation}
    \begin{aligned}
            &\delta=\qty(0^{15},1): D_{16} \quad  &&\delta=\qty(0^{12},\qty(\frac{1}{2})^4):D_4\times D_{12}  ,\\
        &\delta=\qty(\qty(\frac{1}{4})^{16}):A_{15}\times U(1)  ,
            &&\delta=\qty(\qty(\frac{1}{2})^8,0^8):D_8\times D_8  ,\\    
    \end{aligned}
\end{equation}
where the last one corresponds to $\delta^2=2$. We have also listed the unbroken gauge subalgebra of $\Spin(32)/\Z_2$ after the orbifold.
For $E_8\times E_8$ the shifts are
\begin{equation}
    \begin{aligned}
        &\delta=(0^8;0^7,1):  &&D_8\times E_8\\
&\delta=\qty(0^6,\frac12,\frac12;0^6,\frac12,\frac12):&&(E_7\times A_1)^2\\
        &\delta=(1,0^7;1,0^7): &&D_8\times D_8 \, ,
    \end{aligned}
\end{equation}
where the last one corresponds to $\delta^2=2$. For each shift we list the unbroken gauge subalgebra of $E_8\times E_8$.

Except for $D_8\times D_8$ strings, the other six heterotic strings have tachyons. This means that the perturbative expansion is around an unstable vacuum. Tachyon condensation of these theories was studied in \cite{hellerman2008stablevacuumtachyonice8,Kaidi_2021}, where it is argued that their (meta)-stable vacuum is in lower dimensions. It was recently shown that these non-supersymmetric heterotic strings describe the dynamics of near-horizon limit of heterotic branes \cite{Kaidi:2023tqo,Kaidi:2024cbx}, which are understood in the context of swampland program in quantum gravity \cite{mcnamara2019cobordismclassesswampland}. Fascinating relations of (non-supersymmetric) heterotic strings  to a kind of generalized cohomology, called \textit{topological modular forms} \cite{Tachikawa:2021mby,Gukov:2018iiq,stolz2011supersymmetric}, were pointed out in \cite{Tachikawa:2023lwf,Saxena:2024eil}.

\begin{figure}[H]
\centering
\begin{tikzpicture}[
  every node/.style={font=\normalsize},
  theory/.style={align=center, inner sep=2pt},
  arr/.style={-{Latex[length=2.4mm]}, line width=0.55pt, shorten <=1pt, shorten >=1pt},
  cftar/.style={{Latex[length=2.1mm]}-{Latex[length=2.1mm]}, line width=0.55pt, shorten <=1pt, shorten >=1pt}
]
  \def\xsep{5.75}
  \def\ysep{2.15}

  \node[theory] (susy10)    at (0,0)             {10d SUSY\\heterotic strings :  $Z^{0,0}\chvec{0}{0}$};
  \node[theory] (nonsusy10) at (\xsep,0)         {10d non-SUSY\\heterotic strings  :  $Z^{2k,2l}\chvec{0}{0}$};

  \node[theory] (k3L)       at (0,-\ysep)        {on K3 :  $Z^{0,0}\chvec{c}{d}$};
  \node[theory] (k3R)       at (\xsep,-\ysep)    {on K3 :  $Z^{2k,2l}\chvec{c}{d}$};

  \node[theory] (enL)       at (0,-2*\ysep)      {on Enriques :  $Z^{k,l}\chvec{c}{d}$};
  \node[theory] (enR)       at (\xsep,-2*\ysep)  {on Enriques :  $Z^{k,l}\chvec{c}{d}$};

  \draw[arr] (susy10) -- node[above, yshift=2pt] {$h^2,\,\mathbb Z_2$} (nonsusy10);
  \draw[arr] (susy10) -- node[left, xshift=-2pt] {$g,\,\mathbb Z_2$} (k3L);
  \draw[arr] (nonsusy10) -- node[right, xshift=2pt] {$g,\,\mathbb Z_2$} (k3R);
  \draw[arr] (k3L) -- node[left, xshift=-2pt] {$h,\,\mathbb Z_4$} (enL);
  \draw[arr] (k3R) -- node[right, xshift=2pt] {$h,\,\mathbb Z_2$} (enR);

  \draw[cftar] (enL) -- node[above, yshift=1pt] {CFT} (enR);
\end{tikzpicture}
\caption{Two equivalent constructions of the heterotic Enriques compactification. On the left side $\Z_2$ orbifold by $g$ and then $\Z_4$ orbifold by $h$ are performed. On the right side, three $\Z_2$ orbifolds in terms of $h^2,g,h$  are carried out.}
\label{fig:heterotic-enriques-square}
\end{figure}
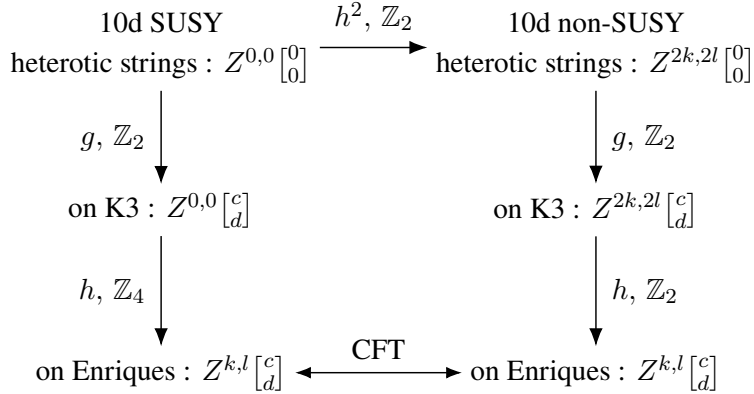

We can associate each Enriques shift vector $w$ satisfying $2w=\delta$ modulo vectors in $\Gamma_{0,16}$ to its "parent" 10-dimensional non-supersymmetric heterotic theory with shift $\delta$. This means that the theories we constructed from supersymmetric strings can be also viewed as compactifications of these non-supersymmetric theories (see figure \ref{fig:heterotic-enriques-square}). Notice, however, that the shift vector $w$ breaks not only the original $E_8\times E_8,\Spin(32)/\Z_2,$ but also part of the residual gauge symmetry of the 10d non-susy heterotic strings. For some shift vectors $w$, we succeeded in removing a part of tachyons in $h^2$-twisted sector, which are coming from parent heterotic strings. For some shift vectors \(w\), part of the tachyonic spectrum in the \(h^2\)-twisted sector, inherited from the parent heterotic strings, is projected out.

\subsection{On spin structure}
Enriques surfaces do not admit spin structures. Indeed, the free part
of the second cohomology lattice of an Enriques surface \(X\) has
intersection form
\begin{equation}
    H^2(X,\mathbb Z)_{\rm free}\simeq E_8(-1)\oplus \Gamma_{1,1},
\end{equation}
and hence signature \(\sigma=-8\).  If \(X\) were spin, Rokhlin's
theorem would imply \(\sigma\in 16\mathbb Z\), which is not the
case.  Therefore Enriques surfaces are non-spin.

The absence of a spin structure on Enriques surfaces may be one of the reasons why string compactifications on them have not been discussed so far in the literature. Nevertheless, superstring theories can be formulated on Enriques surfaces to some extent. For example, taking Enriques orbifold of Type IIA on a K3 surface yields Type 0A string theory on an Enriques surface \cite{Ishige:2026ihp}. This is a natural argument because there are no fermions in Type 0A, at least perturbatively.

A more sophisticated answer will be provided in the case of heterotic strings on a non-spin manifold \cite{Yonekura:2022reu}. In the present work we define the theories as exactly solvable orbifold CFTs. The question of its interpretation as a smooth heterotic compactification on a non-spin Enriques surface, including possible global anomalies and gauge-bundle/torsion data, will be treated in the future.

\section{Light spectrum}\label{sec: light spectrum}

In this section we discuss the tachyonic and massless spectrum of heterotic strings on an Enriques surface. We denote by $p$ the momenta taking values in $\overline \Gamma^{k,l}_{16}\chvec{c}{d}$, $q_{L,R}$ the momenta taking values in $\Gamma^{k,l}_{4,4}\chvec{c}{d}$ and $r$ the $SO(8)$ weights. Since we are interested in the light spectrum, we will analyse the mass formulas for the Neveu-Schwarz (NS) and Ramond (R) states with minimal contribution. To this end, the leading powers of $q$ in $\Theta^{k,l}\chvec{c}{d}$ are summarised in table \ref{tab:leading_contributions}. We will refer to tachyonic states that are present throughout moduli space as \emph{moduli-independent}, whereas tachyons that occur only on special loci in moduli space will instead be called \emph{moduli-dependent}.

Our aim is to show that an appropriate choice of shift vector $w$ can remove moduli-independent tachyons. Moreover, all moduli-dependent tachyons, except the ones coming from the $h^2$-twisted sector, are absent for at least some choice of radii in the compactification space.

\begin{table}[H]
\centering
\small
\renewcommand{\arraystretch}{1.4}
\begin{tabular}{c|c|c}
$g^c h^k$-twisted sector
&
NS leading term
&
R leading term
\\ \hline 

$1$
&
$q^{1/2}$
&
$\, q^{1/2}$
\\[2mm]

$h$
&
$q^{1/8}$
&
$q^{3/8}$
\\[2mm]

$h^2$
&
$q^0$
&
$q^{1/2}$
\\[2mm]

$h^3$
&
$q^{1/8}$
&
$q^{3/8}$
\\ 

$g$
&
$q^{1/4}$
&
$q^{1/4}$
\\[2mm]

$gh$
&
$q^{1/8}$
&
$q^{3/8}$
\\[2mm]

$gh^2$
&
$q^{1/4}$
&
$q^{1/4}$
\\[2mm]

$gh^3$
&
$q^{1/8}$
&
$q^{3/8}$
\end{tabular}

\caption{
Leading contributions in the $q$-expansion of the fermionic partition function $\Theta^{k,l}\chvec{c}{d}$ in the Neveu--Schwarz (NS) and Ramond (R) sectors for each twisted sector $g^c h^k$. Only the lowest powers of $q$ relevant for the light spectrum analysis are displayed.}
\label{tab:leading_contributions}
\end{table}

To make discussions about projection phases more compact in the following, let us define
\begin{align}\label{eq: phases}
    \Phi^{k,l}\chvec{c}{d} =& e^{-\pi \i \,(d v_{1f}+l v_{2f})\cdot(c v_{1f}+k v_{2f})}
e^{\pi \i \,(d v+l w)\cdot(c v+k w)}\\
\times&e^{2\pi \i \,(r+c v_{1f}+k v_{2f})\cdot(d v_{1f}+l v_{2f})}
e^{-2\pi \i \,(p+c v+k w)\cdot(d v+l w)} \, .
\end{align}

\paragraph{Untwisted sector.}
The masses are
\begin{align}
    m^2_{L} = \frac12q_L^2 +\frac12r^2+N_L-\frac12 \, ; \quad m^2_R= \frac12q^2_R + \frac12p^2+N_R-1 \, ,
\end{align}
where the oscillators take integer values. For the lightest NS and R states one finds
\begin{align}
    m^2_L = N_L \, ,
\end{align}
which means that there are no tachyons, although massless states are present.
\paragraph{\texorpdfstring{$g$}{g}-twisted sector.}
The masses are
\begin{align}
    m^2_{L} = \frac12(r+v_{1f})^2+N_L-\frac14 \, ; \quad m^2_R=  \frac12(p+v)^2+N_R-\frac34 \, ,
\end{align}
where the oscillators take half-integer values. Lightest NS and R states satisfy
\begin{align}
    m^2_L = N_L \, ,
\end{align}
so this twisted sector is likewise tachyon-free while containing massless states.
\paragraph{\texorpdfstring{$h$}{h}-twisted sector.}
The masses are
\begin{align}
    m^2_{L} = \frac12 q^2_L+\frac12(r+v_{2f})^2+N_L-\frac14 \, ; \quad m^2_R=  \frac12 q^2_R+\frac12(p+w)^2+N_R-\frac34 \, ,
\end{align}
where the oscillators take half-integer values. The minimal contributions in NS and R sectors have mass
\begin{align}\label{eq: h-twisted mass}
    m^2_L(\text{NS}) = \frac12 q^2_L+N_L-\frac18 \, ; \quad m^2_L(\text{R})=\frac12 q^2_L + N_L +\frac18 \, .
\end{align}
Therefore in the R sector there are no massless nor tachyonic states. Moreover, since in the $h$-twisted sector the winding along $R_9$ is shifted (see Appendix \ref{sec:Twisted bosonic fock space}), $q^2=0$ is not a point in the lattice. Consequently, the NS sector may contain moduli-dependent tachyons if
\begin{align}
    q^2_L < \frac14 \, ,
\end{align}
but not moduli-independent ones. Equivalent results apply to the $h^3$-twisted sector.

\paragraph{\texorpdfstring{$gh$}{gh}-twisted sector.}
The masses are
\begin{align}
    m^2_{L} = \frac12 q^2_L+\frac12(r+v_{1f}+v_{2f})^2+N_L-\frac14 \, ; \quad m^2_R=  \frac12 q^2_R+\frac12(p+v+w)^2+N_R-\frac34 \, ,
\end{align}
where the oscillators take half-integer values. The lightest states in NS and R sectors have the same mass as the ones in the $h$-twisted sector (\ref{eq: h-twisted mass}). In this sector the winding along $R_7$ is shifted which means that again there are moduli-dependent tachyons and no massless states.  The same results hold in the $gh^3$-twisted sector.
\paragraph{\texorpdfstring{$h^2$}{h^2}-twisted sector.}
The masses are
\begin{align}
    m^2_{L} = \frac12 q^2_L+\frac12(r+2v_{2f})^2+N_L-\frac12 \, ; \quad m^2_R=  \frac12 q^2_R+\frac12(p+2w)^2+N_R-1 \, ,
\end{align}
where the oscillators take integer values. Lightest NS and R states satisfy
\begin{align}
    m^2_{L}(\text{NS})=\frac12q^2_L+N_L-\frac12 \, ; \quad m^2_L(\text{R}) = \frac12 q^2_L + N_L \, .
\end{align}
In the R sector there are massless states and no tachyonic states. In the NS sector, in addition to moduli-dependent tachyons, which can appear if
\begin{align}
  q^2_L < 1 \, ,
\end{align}
there is also a moduli-independent tachyon with mass $m^2=-\frac12$. Level-matching imposes $(p+2w)^2=1$.
\paragraph{\texorpdfstring{$gh^2$}{gh^2}-twisted sector.}
The masses are
\begin{align}
    m^2_{L} = \frac12(r+v_{1f}+2v_{2f})^2+N_L-\frac14 \, ; \quad m^2_R=  \frac12(p+v+2w)^2+N_R-\frac34 \, ,
\end{align}
where the oscillators take half-integer values. Lightest NS and R states satisfy
\begin{align}
    m^2_L = N_L \, ,
\end{align}
so this twisted sector is tachyon-free while containing massless states.

\subsection{Tachyons}

\paragraph{Moduli-independent tachyons.} Let us start by determining which Enriques shift vectors $w$ remove tachyonic states which are present independently of the value of radii $R_6,\cdots,R_9$. As discussed above, the only sector in which
moduli-independent tachyons can occur is the $h^2$-twisted sector. In this sector
the lightest NS states have

\begin{equation}
    m_L^2({\rm NS})
=
\frac12 q_L^2+N_L-\frac12 ,
\qquad
m_R^2
=
\frac12 q_R^2+\frac12(p+2w)^2+N_R-1 .
\end{equation}
Thus a moduli-independent tachyon has
\begin{equation}
    q_L=q_R=0,\qquad N_L=N_R=0,
\qquad m^2=-\frac12 ,
\end{equation}
and level matching imposes
\begin{equation}
(p+2w)^2=1,
\qquad p\in \Gamma_{0,16}.
\label{eq:h2-tachyon-level-matching}
\end{equation}
It remains to impose the orbifold projection. 

For a candidate tachyon in the $h^2$-twisted sector, namely
$c=0$, $k=2$, the leading NS contribution of the fermionic block carries
the phase
\begin{equation}
    \exp\!\left[-\frac{\pi i}{2}(d+l)\right].
\end{equation}
Therefore the projection factor for a state labelled by $p$ is 
\begin{align}
&
\frac18 \sum_{d=0}^1\sum_{l=0}^3 \Phi^{2,l}\chvec{0}{d}\\
=&
\frac18
\sum_{d=0}^{1}
\sum_{l=0}^{3}
\exp\!\left[
-2\pi i d\left(p\cdot v+\frac12\right)
\right]
\exp\!\left[
-2\pi i l\left(p\cdot w+w^2+\frac14\right)
\right],
\label{eq:h2-tachyon-projection}
\end{align}
where in the second line we used $v\cdot w=\frac14$, as is the case for
the representatives listed in Section \ref{sec: enriques shift vectors}.
Hence a level-matched candidate survives the orbifold projection if and
only if
\begin{equation}
(p+2w)^2=1,
\qquad
2p\cdot v \equiv 1 \pmod 2,
\qquad
p\cdot w+w^2+\frac14\in\mathbb Z .
\label{eq:tachyon-survival-conditions}
\end{equation}
The last condition can be rewritten in a useful $w$-independent form.
Indeed, using \eqref{eq:h2-tachyon-level-matching} and $p^2\in 2\mathbb Z$,
we have
\begin{equation}
p\cdot w+w^2+\frac14
=
\frac{2-p^2}{4}.
\end{equation}
Thus the tachyon survives precisely when
\begin{equation}
(p+2w)^2=1,
\qquad
2p\cdot v \equiv 1 \pmod 2,
\qquad
p^2\equiv 2 \pmod 4 .
\label{eq:tachyon-survival-congruences}
\end{equation}
We now apply this criterion to all admissible Enriques shift vectors.
There are two possible mechanisms by which the moduli-independent tachyon is
removed. First, the set of level-matched candidates
\begin{equation}
\left\{
p\in\Gamma_{0,16}
\ \middle|\
(p+2w)^2=1
\right\},
\end{equation}
may be empty. Second, there are level-matched candidates, but all the
states may fail the congruence conditions
\eqref{eq:tachyon-survival-congruences} and are therefore projected out.

The moduli-independent-tachyon-free shift classes for the $E_8\times E_8$ lattice are summarized in table \ref{tab:E8E8_shifts} and the ones for the $\Spin(32)/\Z_2$ lattice in table \ref{tab:so32_shift}. Thus, $11$ of the $24$  $E_8\times E_8$ shifts and 9 of the 24 $\Spin(32)/\Z_2$ shifts are free of moduli-independent tachyons.

\begin{table}[H]
    \centering
    \[
\renewcommand{\arraystretch}{1.35}
\begin{array}{c|l}
\text{mechanism} & \text{shift classes} \\ \hline
(p+2w)^2=1\ \text{has no solutions}
&
(\mathsf A_4,\mathsf B_5),\ (\mathsf A_5,\mathsf B_5),\ (\mathsf A_6,\mathsf B_1),\ (\mathsf A_6,\mathsf B_2)
\\[1mm]
\text{all level-matched candidates are projected out}
&
(\mathsf A_2,\mathsf B_2),\
\{\mathsf A_7,\mathsf A_8\}\times\{\mathsf B_7,\mathsf B_8\},
(\mathsf A_9,\mathsf B_9),\ (\mathsf A_{10},\mathsf B_{10})
\end{array}
\]
    \caption{Shift vectors for $E_8\times E_8$ without moduli-independent tachyons}
    \label{tab:E8E8_shifts}
\end{table}

\begin{table}[h]
    \centering
    \[
\renewcommand{\arraystretch}{1.35}
\begin{array}{c|l}
\text{mechanism} & \text{shift classes} \\ \hline
(p+2w)^2=1\ \text{has no solutions}
&
W_{0,\frac12}(8,0),\
W_{0,\frac12}(8,2),\
W_{\frac14,\frac14}(6,1),\
W_{\frac14,\frac14}(6,3)
\\[1mm]
\text{all level-matched candidates are projected out}
&
W_{\frac14,\frac14}(2,2),\
W_{\frac14,\frac14}(2,6),\
W_{\frac14,\frac14}(10,0),\
W_{\frac18,\frac38}(0),\
W_{\frac18,\frac38}(1)
\end{array}
\]
    \caption{Shift vectors for $\Spin(32)/\Z_2$ without moduli-independent tachyons.}
    \label{tab:so32_shift}
\end{table}

\paragraph{Moduli-dependent tachyons.}
Moduli-dependent tachyons can occur in the $h$, $gh$, and $h^2$ twisted
sectors.  We use the convention
\begin{equation}
  q_{L,i}
  =
  {\frac{1}{\sqrt{2}}}
  \left(
    {\frac{n_i}{R_i}}+m_i R_i
  \right),
  \qquad
  q_{R,i}
  =
  {\frac{1}{\sqrt{2}}}
  \left(
    {\frac{n_i}{R_i}}-m_i R_i
  \right),
  \label{eq:qLqR-rectangular}
\end{equation}
where \(i=6,\ldots,9\).  

In the $h$-twisted sector the half-shifted
winding lies along the $R_9$ circle. The half-shifted winding implies that $q^2_L-q^2_R \in \mathbb{Z}$. The lightest NS states are tachyonic
when
\begin{equation}
  q_L^2 < {\frac14},
\end{equation}
and level matching gives two possible gauge-lattice branches,
\begin{equation}
  (p+w)^2={\frac54},
  \qquad
  (p+w)^2={\frac14}.
  \label{eq:h-tachyon-branches}
\end{equation}
For the branch \((p+w)^2=5/4\), the lightest state has
\((m_9,n_9)=(1/2,0)\), and hence
\begin{equation}
  q_L^2 = {\frac{R_9^2}{8}}.
\end{equation}
Thus this branch is tachyonic for
\begin{equation}
  R_9^2 < 2.
  \label{eq:h-tachyon-5over4-radius}
\end{equation}
For the branch \((p+w)^2=1/4\), the lightest choice is
\((m_9,n_9)=(1/2,-1)\), for which
\begin{equation}
  q_L^2
  =
  {\frac12}
  \left(
    {\frac{R_9}{2}}-{\frac{1}{R_9}}
  \right)^2 .
\end{equation}
This state is tachyonic for
\begin{equation}
  3-\sqrt{5}
  <
  R_9^2
  <
  3+\sqrt{5}.
  \label{eq:h-tachyon-1over4-radius}
\end{equation}
The $h^3$ sector is obtained by replacing \(w\) with \(3w\), and gives the
same radius conditions.

Similarly, in the $gh$-twisted sector the half-shifted winding lies along
the $R_7$ circle.  The two gauge-lattice branches are
\begin{equation}
  (p+v+w)^2={\frac54},
  \qquad
  (p+v+w)^2={\frac14},
  \label{eq:gh-tachyon-branches}
\end{equation}
and the corresponding lightest tachyonic loci are
\begin{equation}
  R_7^2 < 2,
  \qquad
  3-\sqrt{5}
  <
  R_7^2
  <
  3+\sqrt{5}.
  \label{eq:gh-tachyon-radius}
\end{equation}
The $gh^3$ sector is again obtained by replacing \(w\) with \(3w\).

The $h^2$-twisted sector is slightly different. Since no winding is shifted we have $q^2_L-q^2_R \in 2 \mathbb{Z}$. The moduli-dependent
tachyon condition is
\begin{equation}
  (p+2w)^2 = 1,
  \qquad
  q_L^2=q_R^2<1,
  \qquad
  q\neq 0.
  \label{eq:h2-moduli-dependent-tachyon}
\end{equation}
The lightest moduli-dependent tachyons are those with a single entry among $( n_i,m_i)$ equal to $\pm1$. Consider for instance $n_i=1$ for some $i$ and all windings set to zero, leading to
\begin{align}
    q^2_L =\frac{1}{2R_i^2} <1 \, .
\end{align}
So the tachyon condition is\footnote{These tachyons can be interpreted as KK modes of parent tachyons compactified on Enriques surfaces. They are now, however, taking their values in a section of a vector bundle on Enriques surfaces, which is twisted along a loop generating the fundamental group $\pi_1(\text{Enriques})=\Z_2$. }
\begin{align}\label{eq: tach h2}
    R_i^2 >\frac12 \, .
\end{align}
However, for each such configuration there is a corresponding one with all momenta $n$ set to zero and $m_i=1$, leading to
\begin{align}
    q^2_L = \frac12 R^2_i <1 \, .
\end{align}
Thus, this configuration is tachyonic if
\begin{align}
    R^2_i <2 \, ,
\end{align}
and this condition combined with \eqref{eq: tach h2} covers all possible values of $R_i$.
It is clear then that we can deform the compactification space in such a way that we get rid of moduli-dependent tachyons coming from $h$ and $gh$-twisted sectors, but we cannot get rid of all tachyons coming from the $h^2$-twisted sector in general. The only theories with no tachyons are then the ones which do not admit any solution to $(p+2w)^2=1$.

\begin{table}[H]
\centering
\small
\renewcommand{\arraystretch}{1.16}
\setlength{\tabcolsep}{6pt}
\begin{tabular}{c c c c}
\toprule
Sector & Mode & Nature & Tachyonic region \\
\midrule
$1,\ g,\ gh^2$ & -- & none & -- \\
\addlinespace[2pt]
$h,\ h^3$
& $(n_9,m_9)=(0,\pm\frac12)$
&  semi-loc.
& $R_9^2<2$ \\
$h,\ h^3$
& $(n_9,m_9)=(\mp1,\pm\frac12)$
& semi-loc.
& $3-\sqrt5<R_9^2<3+\sqrt5$ \\
\addlinespace[2pt]
$gh,\ gh^3$
& $(n_7,m_7)=(0,\pm\frac12)$
& semi-loc.
& $R_7^2<2$ \\
$gh,\ gh^3$
& $(n_7,m_7)=(\mp1,\pm\frac12)$
&  semi-loc.
& $3-\sqrt5<R_7^2<3+\sqrt5$ \\
\addlinespace[2pt]
$h^2$
& $q_L=q_R=0$
& bulk$^\dagger$
& all $R_i$ \\
$h^2$
& $(n_i,m_i)=(\pm1,0)$
&  bulk
& $R_i^2>\frac12$ \\
$h^2$
& $(n_i,m_i)=(0,\pm1)$
&  bulk
& $R_i^2<2$ \\
\bottomrule
\end{tabular}
\caption{Lightest tachyon candidates by sector.    ``semi-loc.'' means localized in the
reflected directions by $h,h^3,gh,gh^3$.  In the $h^2$ rows,
$i=6,\ldots,9$.  The dagger indicates that the $h^2$ moduli-independent
candidate is present only if it survives the gauge-lattice projection.}
\label{tab:lightest-tachyon-windows}
\end{table}

\subsection{Massless states}
In this section we analyse the massless spectrum of heterotic strings on Enriques. We restrict to solutions with $q_{L,R}^2=0$. At special loci in the moduli space of admissible $\Gamma_{4,4}$ lattices the number of massless states might be enhanced from non-vanishing values of $q_{L,R}$; these states will be disregarded in the following.

Massless states are organised into representations of the six dimensional little group $\Spin(4) \sim SU(2) \times SU(2)$. Representations of $SU(2) \times SU(2)$ will be denoted $(\bullet,\bullet)$, where each entry gives the dimension of the corresponding $SU(2)$ representation. 

As an example, consider the states associated to $N_L=1$ in heterotic compactified on $T^4/G$. In 10 dimensions the right-moving excitations comprise 16 internal oscillators in the trivial representations of $SO(8)$ and 8 spacetime oscillators in the vector $\mathbf{8_v}$ of $SO(8)$. Upon branching
\begin{align}
    SO(8) \longrightarrow SO(4)\simeq SU(2)\times SU(2) \, ,
\end{align}
the  six-dimensional states organise as $(\mathbf{2},\mathbf{2})+20(\mathbf{1},\mathbf{1})$.

The computation of massless spectrum proceeds in three steps. First, we solve the level-matched conditions derived at the beginning of this section. We then branch the relevant $SO(8)$ representations to $SU(2)\times SU(2)$. Finally, multiplicities are determined from the phases appearing in the partition function (\ref{eq: enriques pf pieces}), which have been collected in the function $\Phi^{k,l}\chvec{c}{d}$ \eqref{eq: phases}.

We will use the following relations: $v_{1f}\cdot v_{2f}=\frac14,\,\, v_{1f}^2=\frac12, \, \, v_{2f}^2=\frac14$  and $v\cdot w=\frac14$.
\paragraph{Untwisted sector.} 
The analysis splits into two classes of level-matched solutions
\begin{align}
    r^2=1, q^2_{L}=0=N_L, p^2=0=q^2_R, N_R=1 \, ; \quad r^2=1, q^2_{L}=0=N_L, N_R=0=q^2_R, p^2=2 ,\ .
\end{align}
When $p^2=0,r^2=1$ the surviving states are
\begin{align}\label{eq: sol1 untw}
    (\mathbf{3},\mathbf{3})+(\mathbf{3},\mathbf{1})+(\mathbf{1},\mathbf{3})+16 (\mathbf{2},\mathbf{2})+9(\mathbf{1},\mathbf{1}) \, ,
\end{align}
which correspond to the graviton, B field, dilaton, 16 neutral gauge bosons and 8 neutral scalars.
The second solution explicitly depends on the values of dot products between the 480 roots of $E_8\times E_8$ or $\Spin(32)/\Z_2$ and the shift vectors. We get
\begin{align}   u_1(\mathbf{2},\mathbf{2})+u_2(\mathbf{1},\mathbf{1}) + u_3(\mathbf{1},\mathbf{2})+u_4(\mathbf{2},\mathbf{1}) 
\end{align} 
states, where values of $u_{1,2,3,4}$ for each shift vector are summarised in tables \ref{tab: massless e8e8} and \ref{tab: massless so32}.

\paragraph{\texorpdfstring{$g$}{g}-twisted sector.} There are two solutions to the mass constraints
\begin{align}
    (r+v_{1f})^2=\frac12, N_L=0=N_R, (p+v)^2=\frac32\, ; \quad (r+v_{1f})^2=\frac12,N_L=0, (p+v)^2=\frac12=N_R \, ,
\end{align}
which behave differently under the orbifold phases. While the second is completely projected out, the first one contributes non-trivially.
Taking into account the fixed point degeneracy of 16, it yields
\begin{align}
    \frac{16}{8}\sum_{l=0}^3\sum_{d=0}^1\Phi^{0,l}\chvec{1}{d} =16\cdot \frac{1+h^2}{4}=\begin{cases}
         4+4e^{-4 \pi \i (p\cdot w)} \quad \text{NS}\\
         4-4e^{-4 \pi \i (p\cdot w)} \quad \text{R}
    \end{cases} \, ,
\end{align}
where the lightest eigenstates have eigenvalue of $g$ equal to 1, so the projection factor $\frac{1+g}{2}$ simplifies. Moreover, the insertion of $h,h^3$ in $g$-twisted sector vanishes, as derived in Appendix \ref{sec:Twisted bosonic fock space} : $\Gamma_{4,4}^{0,l}\chvec{1}{d}=0$ for $l=1,3$.
Supposing there are $k_1$ solutions to $(p+v)^2=\frac32$ such that $p\cdot4w$ is even and $k_2$ such that it is odd, the resulting spectrum is
\begin{align}
    16 k_1(\mathbf{1},\mathbf{1}) + 8k_2(\mathbf{2},\mathbf{1}) \, ,
\end{align}
where the possible values of $k_{1,2}$ can be found in tables \ref{tab: massless e8e8} and \ref{tab: massless so32}.
\paragraph{\texorpdfstring{$h^2$}{h^2}-twisted sector.}There is a single level-matched solution
\begin{align}
    (r+2v_{2f})^2 =1, N_L=0=q^2_L, (p+2w)^2=2, N_{R}=0=q^2_R \, ,
\end{align}
corresponding to $2(\mathbf{1},\mathbf{2})+2(\mathbf{2},\mathbf{1})$. Supposing that there are
\begin{itemize}
    \item $m_1$ solutions to $(p+2w)^2=2$ such that $2p\cdot w=\frac12$ mod 2 and $2p \cdot v= 2 \mathbb{Z}$;
    \item $m_2$ solutions to $(p+2w)^2=2$ such that $2p\cdot w=\frac12$ mod 2 and $2p \cdot v= 2 \mathbb{Z}+1$;
    \item $m_3$ solutions to $(p+2w)^2=2$ such that $2p\cdot w=\frac32$ mod 2 and $2p \cdot v= 2 \mathbb{Z}$;
    \item $m_4$ solutions to $(p+2w)^2=2$ such that $2p\cdot w=\frac32$ mod 2 and $2p \cdot v= 2 \mathbb{Z}+1$,
\end{itemize}
the multiplicities of states are
\begin{align}
    \frac18\sum_{l=0}^3\sum_{d=0}^1\Phi^{2,l}\chvec{0}{d} =\begin{cases}
        \frac12  e^{-4 \pi \i  w^2} (m_1+m_3) \left(-1+e^{4 \pi \i w^2} \right) \quad \text{for} \, \, (\mathbf{1},\mathbf{2})\\
        \frac12 e^{-4 \pi \i  w^2} (m_2+m_4) \left(-1+e^{4 \pi \i w^2} \right) \quad \text{for} \, \, (\mathbf{2},\mathbf{1})
    \end{cases} \quad .
\end{align}
The shift vectors $w$ on $E_8\times E_8$ and $\Spin(32)/\Z_2$ fall into two categories: $w^2=\frac14$ mod 1 and $w^2=\frac34$ mod 1. Substituting the values, we find
\begin{align}
        (m_1+m_3)(\mathbf{1},\mathbf{2})+(m_2+m_4)(\mathbf{2},\mathbf{1}) \, ,
\end{align}
for both classes of shift vectors. Values of $m_{1,2,3,4}$ for each choice of shift vector can be found in tables \ref{tab: massless e8e8} and \ref{tab: massless so32}.
\paragraph{\texorpdfstring{$gh^2$}{gh^2}-twisted sector.} Similarly to what happens in the $g$-twisted sector we have
\begin{align}
    (r+v_{1f}+2v_{2f})^2=\frac12, N_L=0=N_R, (p+v+2w)^2=\frac32\, ; \nonumber \\
    (r+v_{1f}+2v_{2f})^2=\frac12,N_L=0, (p+v+2w)^2=\frac12=N_R \, .
\end{align}
Again the second class of solutions is projected out and the first one is non-trivial, leading to
\begin{align}
    \frac{16}{8}\sum_{l=0}^3\sum_{d=0}^1\Phi^{2,l}\chvec{1}{d} =\begin{cases}
         4-4e^{-2 \pi \i p\cdot v} \quad \text{NS}\\
         4+4e^{-2 \pi \i p\cdot v} \quad \text{R}
    \end{cases} \, ,
\end{align}
where we have taken into account the fixed point degeneracy of 16.
Supposing there are $n_1$ solutions to $(p+v+2w)^2=\frac32$ such that $p\cdot2v$ is odd and $n_2$ such that it is even, we get
\begin{align}
    16 n_1(\mathbf{1},\mathbf{1}) + 8n_2(\mathbf{1},\mathbf{2}) \, .
\end{align}
Admissible values of $n_{1,2}$ are listed in tables \ref{tab: massless e8e8} and \ref{tab: massless so32}.
As expected, since $h^2$ acts trivially on bosons, the NS spectrum of this twisted sector is the same as the one in the $g$-sector, while states in the R sector have opposite chirality.

\begin{table*}[ht]
\centering
\scriptsize
\setlength{\tabcolsep}{2.4pt}
\renewcommand{\arraystretch}{1.18}
\resizebox{\textwidth}{!}{%
\begin{tabular}{llccccccccccc}
\toprule
Shift pairs
& 10d parent
& $k_1$ & $k_2$ & $n_1$ & $n_2$
& $m_1$ & $m_2$ & $m_3$ & $m_4$
& $u_2$ & $u_3$ & $u_4$ \\
\midrule
$\{\mathsf A_i,\mathsf B_j\},\; i=1,2,\; j=1,2$
& $E_8\times D_8$
& 56 & 0 & 56 & 0 & 0 & 0 & 0 & 0
& 96 & 64 & 64 \\

$\{\mathsf A_3,\mathsf B_j\},\; j=1,2$
& $E_8\times D_8$
& 24 & 32 & 24 & 32 & 32 & 32 & 32 & 32
& 96 & 64 & 64 \\

$\{\mathsf A_i,\mathsf B_j\},\; i=4,5,\; j=3,4$
& $D_8\times E_8$
& 56 & 0 & 32 & 0 & 0 & 64 & 0 & 64
& 224 & 128 & 0 \\

$\{\mathsf A_i,\mathsf B_5\},\; i=4,5$
& $D_8\times D_8$
& 24 & 32 & 0 & 32 & 32 & 96 & 32 & 96
& 96 & 192 & 64 \\

$\{\mathsf A_6,\mathsf B_j\},\; j=1,2$
& $D_8\times D_8$
& 24 & 32 & 0 & 32 & 32 & 96 & 32 & 96
& 96 & 192 & 64 \\

$\{\mathsf A_i,\mathsf B_6\},\; i=7,8$
& $(E_7\times A_1)^2$
& 32 & 24 & 8 & 24 & 24 & 88 & 24 & 88
& 128 & 176 & 48 \\

$\{\mathsf A_i,\mathsf B_j\},\; i=7,8,\; j=7,8$
& $(E_7\times A_1)^2$
& 0 & 56 & 0 & 56 & 56 & 56 & 56 & 56
& 0 & 112 & 112 \\

$\{\mathsf A_i,\mathsf B_j\},\; i=9,10,\; j=9,10$
& $(E_7\times A_1)^2$
& 32 & 24 & 8 & 24 & 24 & 88 & 24 & 88
& 128 & 176 & 48 \\

\bottomrule
\end{tabular}
}
\caption{Grouped massless state multiplicities for the $E_8\times E_8$
models. The \(u_i\) columns refer to the untwisted charged \(p^2=2\) sector, while \(k_i,n_i,m_i\) denote twisted-sector multiplicities.}
\label{tab: massless e8e8}
\end{table*}
\begin{table*}[ht]
\centering
\scriptsize
\setlength{\tabcolsep}{2.6pt}
\renewcommand{\arraystretch}{1.18}
\resizebox{\textwidth}{!}{%
\begin{tabular}{@{}llllrrrrrrrrrrr@{}}
\toprule
shift family & label & 10d parent & gauge algebra
& \(k_1\) & \(k_2\)
& \(n_1\) & \(n_2\)
& \(m_1\) & \(m_2\) & \(m_3\) & \(m_4\)
& \(u_2\) & \(u_3\) & \(u_4\) \\
\midrule
\(W_{0,\frac12}(0,n)\)
& \(n=0,2,4,6\)
& \(D_{16}\)
& \(D_{14-n}\times D_n\times U(1)^2\)
& 56 & 0 & 56 & 0 & 0 & 0 & 0 & 0 & 224 & 0 & 0 \\

\(W_{0,\frac12}(4,n)\)
& \(n=1,3,5\)
& \(D_4\times D_{12}\)
& \(D_{10-n}\times A_3\times D_n\times U(1)^3\)
& 40 & 16 & 16 & 16 & 16 & 80 & 16 & 80 & 160 & 160 & 32 \\

\(W_{0,\frac12}(12,1)\)
& --
& \(D_4\times D_{12}\)
& \(A_{11}\times U(1)^5\)
& 8 & 48 & 8 & 48 & 48 & 48 & 48 & 48 & 32 & 96 & 96 \\

\(W_{\frac14,\frac14}(2,n)\)
& \(n=0,2,4,6\)
& \(D_4\times D_{12}\)
& \(D_{12-n}\times A_1^2\times D_n\times U(1)^2\)
& 8 & 48 & 8 & 48 & 48 & 48 & 48 & 48 & 32 & 96 & 96 \\

\(W_{\frac14,\frac14}(10,n)\)
& \(n=0,2\)
& \(D_4\times D_{12}\)
& \(D_{4-n}\times A_9\times D_n\times A_1\times U(1)^2\)
& 40 & 16 & 16 & 16 & 16 & 80 & 16 & 80 & 160 & 160 & 32 \\

\(W_{0,\frac12}(8,n)\)
& \(n=0,2\)
& \(D_8\times D_8\)
& \(D_{6-n}\times A_7\times D_n\times U(1)^3\)
& 24 & 32 & 0 & 32 & 32 & 96 & 32 & 96 & 96 & 192 & 64 \\

\(W_{\frac14,\frac14}(6,n)\)
& \(n=1,3\)
& \(D_8\times D_8\)
& \(D_{8-n}\times A_5\times A_1\times D_n\times U(1)^2\)
& 24 & 32 & 0 & 32 & 32 & 96 & 32 & 96 & 96 & 192 & 64 \\

\(\widetilde W_{\frac14,\frac14}(0)\)
& --
& \(D_4\times D_{12}\)
& \(D_{12}\times A_1\times A_1\times U(1)^2\)
& 8 & 48 & 8 & 48 & 48 & 48 & 48 & 48 & 32 & 96 & 96 \\

\(\widetilde W_{\frac14,\frac14}(1)\)
& --
& \(D_4\times D_{12}\)
& \(D_4\times A_9\times A_1\times U(1)^2\)
& 40 & 16 & 16 & 16 & 16 & 80 & 16 & 80 & 160 & 160 & 32 \\

\(W_{\frac18,\frac38}(n)\)
& \(n=0,1\)
& \(A_{15}\times U(1)\)
& \(A_{10-4n}\times A_{2+4n}\times U(1)^4\)
& 28 & 28 & 4 & 28 & 28 & 92 & 28 & 92 & 112 & 184 & 56 \\

\(W_{\frac58,-\frac18}(n)\)
& \(n=0,1\)
& \(A_{15}\times U(1)\)
& \(A_{12-4n}\times A_{4n}\times U(1)^4\)
& 28 & 28 & 4 & 28 & 28 & 92 & 28 & 92 & 112 & 184 & 56 \\
\bottomrule
\end{tabular}
}
\caption{
Grouped massless-state multiplicities for the \(\mathrm{Spin}(32)/\mathbb Z_2\) Enriques models. The table combines the shift family, ten-dimensional non-supersymmetric parent,
and unbroken gauge algebra. }
\label{tab: massless so32}
\end{table*}

The massless spectrum content is summarised in tables \ref{tab: massless e8e8} and \ref{tab: massless so32}.
The quantity $u_1$ has been omitted from the tables as it can be reconstructed as the dimension of the adjoint representation of the unbroken gauge algebra (see \eqref{eq:list_of_AB} for $E_8 \times E_8$) minus 16. An especially interesting feature of the $E_8 \times E_8$ models is that four of the allowed shift vectors project out all twisted sector scalars, thereby yielding compactifications with a reduced moduli content. Recently, compactifications with no moduli at all except the dilaton, called islands, have drawn attention \cite{Aldazabal:2025zht,Baykara:2025lhl}.

\section*{Acknowledgements}
We are grateful to Ida Zadeh for helpful discussions during the early stages of this work and for useful comments on an earlier version of the manuscript. We also thank Yuta Hamada, Hiroki Wada, and Masashi Kawahira for useful comments and discussions. A.I. also thanks Shun'ya Mizoguchi for his encouragement throughout this work.

\begin{appendix}
    
\section{Theta functions}\label{sec:theta functions}
The theta functions are

\begin{equation}\label{eq:thetas}
\begin{aligned}
\vartheta_1=\vartheta\Bigl[\begin{matrix}1\\1\end{matrix}\Bigr]\coloneqq &i \sum_{n \in \mathbb{Z}}(-1)^n q^{\frac{1}{2}\qty(n-\frac{1}{2})^2}=0, \\
\vartheta_2=\vartheta\Bigl[\begin{matrix}1\\0\end{matrix}\Bigr]\coloneqq&\sum_{n \in \mathbb{Z}} q^{\frac{1}{2}\qty(n-\frac{1}{2})^2}  =2 q^{\frac{1}{8}} \prod_{n=1}^{\infty}\qty(1-q^n)\qty(1+ q^n)\qty(1+ q^n),\\
\vartheta_3=\vartheta\Bigl[\begin{matrix}0\\0\end{matrix}\Bigr]\coloneqq&\sum_{n \in \mathbb{Z}} q^{\frac{1}{2} n^2} =\prod_{n=1}^{\infty}\qty(1-q^n)\qty(1+ q^{n-\frac{1}{2}})\qty(1+ q^{n-\frac{1}{2}}),\\
\vartheta_4=\vartheta\Bigl[\begin{matrix}0\\1\end{matrix}\Bigr]\coloneqq&\sum_{n \in \mathbb{Z}}(-1)^n q^{\frac{1}{2} n^2}=\prod_{n=1}^{\infty}\qty(1-q^n)\qty(1- q^{n-\frac{1}{2}})\qty(1-q^{n-\frac{1}{2}}).
\end{aligned}
\end{equation}
A more general definition is given by

\begin{equation}
    \tchar{a}{b}=\sum_{n\in\Z}q^{\frac{1}{2}\qty(n+\frac{a}{2})^2}e^{\pi i b\qty(n+\frac{a}{2})}, a,b\in\Z.
\end{equation}
Sometimes we will use the short-hand notation, e.g. $\vartheta_{10}$ for $\vartheta\chvec{1}{0}$.

We used the following equations elsewhere:

\begin{equation}\label{eq:theta_priodicity}
\begin{aligned}
\vartheta\chvec{a\pm 2}{b}&=\vartheta\chvec{a}{b},&&\vartheta\chvec{a}{b\pm 2}=e^{\pm\pi i a}\vartheta\chvec{a}{b},\\
T\cdot\vartheta\chvec{a}{b}=&e^{-\frac{1}{4}\pi i a(a-2)}\vartheta\chvec{a}{a+b-1},&&S\cdot\vartheta\chvec{a}{b}=\qty(-i\tau)^{\frac{1}{2}}e^{\frac{1}{2}\pi i ab}\vartheta\chvec{b}{-a},\\
\end{aligned}
\end{equation}

\begin{equation}
    \prod_{n=1}^\infty\frac{1}{(1+q^n)^4}=\frac{4\eta^6}{\vartheta_{10}^2}\prod_{n=1}\frac{1}{(1-q^n)^4}.
\end{equation}

The definition of the Dedekind eta function and its transformation are
\begin{equation}
    \begin{aligned}
        &\eta(\tau)=q^{\frac{1}{24}}\prod_{n=1}^\infty \qty(1-q^n),\\
        \eta(\tau+1)&=e^{\frac{\pi i}{12}}\eta(\tau),
        \quad\eta\qty(-\frac{1}{\tau})=\sqrt{-i\tau}\eta(\tau).
    \end{aligned}
\label{eq:eta_modular_tr}\end{equation}

\section{Twisted bosonic Fock space}\label{sec:Twisted bosonic fock space}

In this section we construct $g,h,gh$-twisted Hilbert spaces for $\H_{\Gamma_{4,4}}$ explicitly, and calculate 
\begin{align}
    \Gamma^{k,l}\chvec{c}{d}=\eta^4\bar\eta^4\tr_{\H_{\Gamma_{4,4}}^{g^c h^k}}g^d h^l\q
\end{align}
as its trace. On the compactification of heterotic strings on $T^4$, there is a Hilbert space associated with $\Gamma_{4,4}=H^1(T^4;\Z)\oplus H_1(T^4;\Z)$:
\begin{equation}
\H_{\Gamma_{4,4}}=\H_{\text{Boson}}^{4,4}\otimes\qty(\bigoplus_{p\in\Gamma_{4,4}}\C \ket{p}).
\end{equation}
The actions of $g,h$ on the fields are
\begin{equation}
    \begin{aligned}
        g:
\begin{cases}
X^6 \rightarrow -X^6,\\
X^7 \rightarrow -X^7 ,\\
X^8 \rightarrow -X^8 ,\\
X^9 \rightarrow -X^9, \\
\end{cases}      \quad \quad\quad   h :
\begin{cases}
X^6 \rightarrow -X^6,\\
X^7 \rightarrow -X^7+\pi R_7 ,\\
X^8 \rightarrow X^8 ,\\
X^9 \rightarrow X^9+ \pi R_9. \\
\end{cases}
    \end{aligned}
\end{equation}
The action of $g,h$ on a state $\ket{p}$ are given as follows:
\begin{equation}
    \begin{aligned}
        g\ket{p_6,p_7,p_8,p_9}=&\ket{-p_6,-p_7,-p_8,-p_9},\\
        h\ket{p_6,p_7,p_8,p_9}=&(-1)^{R_7p_7+R_9 p_9}\ket{-p_6,-p_7,p_8,p_9},\\
gh\ket{p_6,p_7,p_8,p_9}=&(-1)^{R_7p_7+R_9 p_9}\ket{p_6,p_7,-p_8,-p_9}.\\
    \end{aligned}
\end{equation}

Now we can compute $\Gamma^{0,l}\chvec{0}{d}$ as follows:

\begin{equation}
    \begin{aligned}
     &\Gamma_{4,4}^{0,0}\chvec{0}{0}=\sum_{p\in\Gamma_{4,4}}q^{\frac{1}{2}p_L^2}\bar q^{\frac{1}{2}p_R^2},
     &&\Gamma_{4,4}^{0,0}\chvec{0}{1}=\frac{16|\eta|^{12}}{|\vartheta_{10}|^4},\\
    &\Gamma^{0,1}_{4,4}\chvec{0}{0}
        =\biggl|\frac{2\eta^3}{\vartheta_{10}}\biggr|^2\vartheta(R_8)\vartheta^{0,1}(R_9),
        &&\Gamma^{0,1}_{4,4}\chvec{0}{1}
        =\biggl|\frac{2\eta^3}{\vartheta_{10}}\biggr|^2\vartheta(R_6)\vartheta^{0,1}(R_7),
    \end{aligned}
\end{equation}
where $\Gamma_{1,1}(R)$ is a $(1,1)$ even self-dual lattice and  $\vartheta(R),\vartheta^{k,l}(R)$ are its (modified) theta function:

    \begin{align}
\Gamma_{1,1}(R)=&\frac{1}{\sqrt{2}}\Bigl\{\left(\frac{n}{R}+mR;\frac{n}{R}-mR\right)|n,m\in\Z\Bigr\},\\
\vartheta(R)=&\sum_{p\in\Gamma_{1,1}(R)}q^{\frac{1}{2}p_L^2}\bar q^{\frac{1}{2}p_R^2},\\
    \theta^{k,l}(R)=&\sum_{n\in\Z}\sum_{m\in \Z+\frac{1}{2}k}(-1)^{ln}q^{\frac{1}{4}\qty(\frac{n}{R}+mR)^2}\bar q^{\frac{1}{4}\qty(\frac{n}{R}-mR)^2}.
\end{align}

\subsection{\texorpdfstring{$g$}{g}-twisted space}
The boundary conditions of $g$-twisted bosons $X^i_g:T^2\to T^4$ are 

  \begin{equation}
    \begin{aligned}
    X_g^{i}(\tau,\sigma+2\pi)=&g\cdot X_g^i(\tau,\sigma)\\
        =&-X_g^i(\tau,\sigma),\quad6\leq i,j\leq 9.
    \end{aligned}
    \end{equation}

Let us denote the Fock space of $X_g^i$ by $\H^{4,4}_{X_g}$. The $g$-twisted Hilbert space $\H_{\Gamma_{4,4}}^{g}$ is given concretely as follows: 
\begin{equation}
\begin{aligned}
\H_{\Gamma_{4,4}}^{g}=&\H^{4,4}_{X_g}\otimes\qty(\bigoplus_{\text{fixed points}}\C\ket{x_6,x_7,x_8,x_9})\\
\end{aligned}
\end{equation}
where all 16 states $\ket{x_6,x_7,x_8,x_9}$ are $g$ invariant, while half pairs of them are $h$ invariant.
Canonical quantization gives

    \begin{equation}
        \begin{aligned}
            [\alpha^i_r,\alpha^j_s]=&[\tilde{\alpha}^i_r,\tilde{\alpha}^j_s]=r\delta^{ij}\delta_{r+s,0},\\
            [\alpha^i_r,\tilde{\alpha}^j_s]=&0, \quad r,s\in\Z+\frac12,\quad6\leq i,j\leq 9,
        \end{aligned}
    \end{equation}
and
    \begin{equation}
        \begin{aligned}
            L_0=&\frac{1}{2}\sum_{i=6}^9\sum_{r=1/2}\qty(\alpha_{-r}^i\alpha_{r}^i+\alpha_r^i\alpha_{-r}^i)\\
            =&\sum_{i=6}^9\sum_{r=1/2}\alpha_{-r}^i\alpha_{r}^i+\frac{1}{12}.
        \end{aligned}
    \end{equation}

Then it holds that

    \begin{equation}\label{eq:gamma01}
        \begin{aligned}
            &\Gamma_{4,4}^{0,0}\chvec{1}{0}=16\biggl|\frac{\eta^{12}}{\vartheta_{01}^4}\biggr|,
        &&\Gamma_{4,4}^{0,0}\chvec{1}{1}=16\biggl|\frac{\eta^{12}}{\vartheta_{00}^4}\biggr|,\\
   & \Gamma_{4,4}^{0,1}\chvec{1}{0}=\Gamma_{4,4}^{0,1}\chvec{1}{1}=
        0.
    \end{aligned}
\end{equation}

\subsection{\texorpdfstring{$h$}{h}-twisted space}

The boundary conditions of $h$-twisted bosons $X^i_h:T^2\to T^4$ are given as follows:
 \begin{equation}
    \begin{aligned}
    X_h^{i}(\sigma_1+2\pi,\sigma_2)=&h\cdot X_h^i(\sigma_1,\sigma_2),i=6,7,8,9,
    \end{aligned}
    \end{equation}
where
 \begin{equation}
             h :
\begin{cases}
X^6 \rightarrow -X^6,\\
X^7 \rightarrow -X^7+\pi R_7  ,\\
X^8 \rightarrow X^8,\\
X^9 \rightarrow X^9+\pi R_9. \\
\end{cases}
 \end{equation}

Let us denote the Fock space of $X_h^i$ by $\H^{4,4}_{X_h}$.  The $h$-twisted Hilbert space $\H_{\Gamma_{4,4}}^{h}$ is given concretely as follows: 

\begin{equation}
\begin{aligned}
\H_{\Gamma_{4,4}}^{h}=&
\H^{4,4}_{X_h}\otimes \left(\bigoplus_{\substack{x^6=0,\pi R_6 \\ x^7=\pm\frac12\pi R_7\\(p_8,p_9)\in\Gamma_{2,2}(R_8,R_9)}}\C\ket{x_6,x_7,p_8,p_9}\right).\\
\end{aligned}
\end{equation}

The actions of $g,h$ are given by

\begin{equation}
\begin{aligned}
    g\ket{x_6,x_7,p_8,p_9}=&\ket{-x_6,-x_7,-p_8,-p_9},\\
    h\ket{x_6,x_7,p_8,p_9}=&(-1)^{R_9p_9}\ket{-x_6,-x_7+\pi R_7,p_8,p_9},\\
    gh\ket{x_6,x_7,p_8,p_9}=&(-1)^{R_9p_9}\ket{x_6,x_7+\pi R_7,-p_8,-p_9}.
\end{aligned}
\end{equation}
The winding number $m_9$ of $X^9$ takes its value in $\Z+\frac12$ since it holds that
\begin{equation}
    \begin{aligned}
        \pi R_9=&X^9(2\pi)-X^9(0)\\
        =&2\pi m_9 R_9.
    \end{aligned}
\end{equation}
Canonical quantization gives

    \begin{equation}
        \begin{aligned}
            [\alpha^i_r,\alpha^j_s]=&[\tilde{\alpha}^i_r,\tilde{\alpha}^j_s]=r\delta^{ij}\delta_{r+s,0},\\
            [\alpha^i_r,\tilde{\alpha}^j_s]=&0, \quad r,s\in\Z+\frac12,i,j=6,7.
        \end{aligned}
    \end{equation}

 \begin{equation}
        \begin{aligned}
            [\alpha^i_n,\alpha^j_m]=&[\tilde{\alpha}^i_n,\tilde{\alpha}^j_m]=n\delta^{ij}\delta_{n+m,0},\\
            [\alpha^i_n,\tilde{\alpha}^j_m]=&0, \quad n,m\in\Z,i,j=8,9.
        \end{aligned}
    \end{equation}

    $L_0$ is given by

    \begin{equation}
        \begin{aligned}
            L_0=&\frac{1}{2}\sum_{i=6}^7\sum_{r=1/2}\qty(\alpha_{-r}^i\alpha_{r}^i+\alpha_r^i\alpha_{-r}^i)+\frac{1}{2}\sum_{i=8}^9\sum_{n=1}\qty(\alpha_{-n}^i\alpha_{n}^i+\alpha_n^i\alpha_{-n}^i)\\
            =&\sum_{i=6}^7\sum_{r=1/2}\alpha_{-r}^i\alpha_{r}^i+\sum_{i=8}^9\sum_{n=1}\alpha_{-n}^i\alpha_{n}^i-\frac{1}{24}.
        \end{aligned}
    \end{equation}
The result is
\begin{equation}
    \begin{aligned}
        \Gamma_{4,4}^{1,0}\chvec{0}{0}
        =&4\frac{\eta^3\bar\eta^3}{|\vartheta_{01}|^2}\vartheta(R_8)\vartheta^{1,0}(R_9),\\
    \Gamma_{4,4}^{1,0}\chvec{0}{1}=& \Gamma_{4,4}^{1,1}\chvec{0}{1}=0,\\
        \Gamma_{4,4}^{1,1}\chvec{0}{0}
        =&4\frac{\eta^3\bar\eta^3}{|\vartheta_{00}|^2}\vartheta(R_8)\vartheta^{1,1}(R_9),\\
    \end{aligned}
\end{equation}

\subsection{\texorpdfstring{$gh$}{gh}-twisted space}

The boundary conditions of $gh$-twisted bosons $X^i_{gh}:T^2\to T^4$ are

\begin{equation}
    \begin{aligned}
    &X_{gh}^{i}(\sigma_1+2\pi,\sigma_2)\\=&gh\cdot X_{gh}^i(\sigma_1,\sigma_2),~~i=6,7,8,9,
    \end{aligned}
    \end{equation}

    where

     \begin{equation}
             gh :
\begin{cases}
X^6 \rightarrow X^6,\\
X^7 \rightarrow X^7 +\pi R_7 ,\\
X^8 \rightarrow -X^8,\\
X^9 \rightarrow -X^9- \pi R_9. \\
\end{cases}
 \end{equation}
 
The winding number $m_7$ of $X^7$ takes its value in $\Z+\frac12$ in the $gh$-twisted sector: 
\begin{equation}
    \begin{aligned}
        \pi R_7=&X^7(2\pi)-X^7(0)\\
        =&2\pi m_7 R_7.
    \end{aligned}
\end{equation}

    After the canonical quantization, we find

\begin{equation}
        \begin{aligned}
            [\alpha^i_n,\alpha^j_m]=&[\tilde{\alpha}^i_n,\tilde{\alpha}^j_m]=n\delta^{ij}\delta_{n+m,0},\\
            [\alpha^i_n,\tilde{\alpha}^j_m]=&0, \quad n,m\in\Z,i,j=6,7,
        \end{aligned}
    \end{equation}

    \begin{equation}
        \begin{aligned}
            [\alpha^i_r,\alpha^j_s]=&[\tilde{\alpha}^i_r,\tilde{\alpha}^j_s]=r\delta^{ij}\delta_{r+s,0},\\
    [\alpha^i_r,\tilde{\alpha}^j_s]=&0, \quad r,s\in\Z+\frac12,i,j=8,9.
        \end{aligned}
    \end{equation}

 and
    \begin{equation}
        \begin{aligned}
            L_0=&\frac{1}{2}\sum_{i=6}^7\sum_{n=1}\qty(\alpha_{-n}^i\alpha_{n}^i+\alpha_n^i\alpha_{-n}^i)+\frac{1}{2}\sum_{i=8}^9\sum_{r=1/2}\qty(\alpha_{-r}^i\alpha_{r}^i+\alpha_r^i\alpha_{-r}^i)\\
            =&\sum_{i=6}^7\sum_{n=1}\alpha_{-n}^i\alpha_{n}^i+\sum_{i=8}^9\sum_{r=1/2}\alpha_{-r}^i\alpha_{r}^i-\frac{1}{24}.
        \end{aligned}
    \end{equation}

Let us denote the Fock space of $X_{gh}^i$ by $\H^{4,4}_{X_{gh}}$.  The $gh$-twisted Hilbert space $\H_{\Gamma_{4,4}}^{gh}$ is given concretely as follows: 

\begin{equation}
\begin{aligned}
\H_{\Gamma_{4,4}}^{gh}=&
\H^{4,4}_{X_{gh}}\otimes \left(\bigoplus_{\substack{ x^8=0,\pi R_8\\x^9=\pm\frac{1}{2}\pi R_9 \\(p_6,p_7)\in\Gamma_{2,2}(R_6,R_7)}}\C\ket{p_6,p_7,x_8,x_9}\right).\\
\end{aligned}
\end{equation}
The actions of $g,h,gh$ are defined as

\begin{equation}
    \begin{aligned}
        g\ket{p_6,p_7,x_8,x_9}=&\ket{-p_6,-p_7,-x_8,-x_9},\\
        h\ket{p_6,p_7,x_8,x_9}=&(-1)^{R_7 p_7}\ket{-p_6,-p_7,x_8,x_9+\pi R_9},\\
    gh\ket{p_6,p_7,x_8,x_9}=&(-1)^{R_7 p_7}\ket{p_6,p_7,-x_8,-x_9+\pi R_9}.\\
    \end{aligned}
\end{equation}
Then the results are

\begin{equation}
    \begin{aligned}
        \Gamma_{4,4}^{1,0}\chvec{1}{0}
        =&4\frac{\eta^3\bar\eta^3}{|\vartheta_{01}|^2}\vartheta(R_6)\vartheta^{1,0}(R_7),\\
    \Gamma_{4,4}^{1,0}\chvec{1}{1}=&\Gamma_{4,4}^{1,1}\chvec{1}{0}=0,\\
      \Gamma_{4,4}^{1,1}\chvec{1}{1}
        =&4\frac{\eta^3\bar\eta^3}{|\vartheta_{00}|^2}\vartheta(R_6)\vartheta^{1,1}(R_7).
    \end{aligned}
\end{equation}

\subsection{Summary}
In this section we obtain the following expression:
\begin{equation}
    \begin{aligned}
    \Gamma^{0,0}\chvec{c}{d}=&\frac{16|\eta|^{12}}{|\vartheta\chvec{1+c}{1+d}|^4},(c,d)\neq (0,0)\\
        \Gamma^{k,l}\chvec{0}{0}=&\frac{4\eta^3\bar\eta^3}{|\vartheta\chvec{k+1}{l+1}|^2}\theta(R_8)\theta^{k,l}(R_9),\\
         \Gamma^{c,d}\chvec{c}{d}=&\frac{4\eta^3\bar\eta^3}{|\vartheta\chvec{1+c}{1+d}|^2}\theta(R_6)\theta(R_7),(c,d)\neq (0,0).\\
    \end{aligned}
\end{equation}
Others are zero:
\begin{equation}
\begin{aligned}
        \Gamma^{0,1}\chvec{1}{d}=&0,\quad\text{($g$-twisted sector)}\\
        \Gamma^{1,l}\chvec{0}{1}=&0,\quad\text{($h$-twisted sector)}\\
        \Gamma^{1,d+1}\chvec{1}{d}=&0,\quad\text{($gh$-twisted sector)}\\
\end{aligned}
\end{equation}
\section{Root lattices}\label{sec:lie algebra}
In this appendix, we summarise features of the root and weight lattices used in this paper. For notation and more details see \cite{Bourbaki:2002}.
We denote the $i$-th standard orthonormal basis vector of $\R^n$ by $\varepsilon_i$.
\subsection{\texorpdfstring{$A_n$}{An} type}
The root system of $A_n$ is
\begin{equation}
\begin{aligned}
    &\ve_j-\ve_k, 
    &&\text{for}\quad j\neq k
\end{aligned}
\end{equation}
The basis is
        \begin{equation}
\begin{aligned}
        &\alpha_i=\varepsilon_i-\varepsilon_{i+1},&&\text{for}\quad 1\leq i \leq n.
\end{aligned}
\end{equation}

\subsection{\texorpdfstring{$D_n$}{Dn} type}
The root system of $D_n$:
\begin{equation}
\begin{aligned}  &\pm\ve_j\pm\ve_k,\pm\ve_j\mp\ve_k,&&\text{for}\quad 1\leq j<k\leq n.
\end{aligned}
\end{equation}
The basis is
\begin{equation}
\begin{aligned}
        &\alpha_i=\ve_i-\ve_{i+1},&&\text{for}\quad 1\leq i\leq n-1, \\
        &\alpha_n=\ve_{n-1}+\ve_n.
\end{aligned}
\end{equation}

\subsection{\texorpdfstring{$E_6$}{E6}}

The root system of $E_6$ is
\begin{equation}
\begin{aligned}
    &\pm\varepsilon_i\pm\varepsilon_j, 
    &&\text{for}\quad1\leq i<j\leq 5,\\
    &\pm\frac{1}{2}\qty(\varepsilon_8-\varepsilon_7-\varepsilon_6+\sum_{i=1}^5 (-1)^{\nu_i}\varepsilon_i),
    &&\text{for}\quad\sum_{i=1}^5 \nu_i\in 2\Z.
\end{aligned}
\end{equation}
The basis is
\begin{equation}
\begin{aligned}
&\alpha_1=\frac{1}{2}\left(\varepsilon_1+\varepsilon_8\right)-\frac{1}{2}\left(\varepsilon_2+\varepsilon_3+\varepsilon_4+\varepsilon_5+\varepsilon_6+\varepsilon_7\right), &&\alpha_2=\varepsilon_1+\varepsilon_2, \\
& \alpha_3=\varepsilon_2-\varepsilon_1, &&\alpha_4=\varepsilon_3-\varepsilon_2, \\
&\alpha_5=\varepsilon_4-\varepsilon_3, &&\alpha_6=\varepsilon_5-\varepsilon_4.
\end{aligned}
\end{equation}

\subsection{\texorpdfstring{$E_7$}{E7}}

The root system of $E_7$ is
\begin{equation}
\begin{aligned}
    &\pm\varepsilon_i\pm\varepsilon_j, 
    &&\text{for}\quad1\leq i<j\leq 6,
    \\
    &\pm(\varepsilon_7-\varepsilon_8),\\
    &\pm\frac{1}{2}\qty(\varepsilon_8-\varepsilon_7-\varepsilon_6+\sum_{i=1}^6 (-1)^{\nu_i}\varepsilon_i),
    &&\text{for}\quad\sum_{i=1}^6 \nu_i\in 2\Z.
\end{aligned}
\end{equation}
The basis is
\begin{equation}
\begin{aligned}
& \alpha_1=\frac{1}{2}\left(\varepsilon_1+\varepsilon_8\right)-\frac{1}{2}\left(\varepsilon_2+\varepsilon_3+\varepsilon_4+\varepsilon_5+\varepsilon_6+\varepsilon_7\right), \\
& \alpha_2=\varepsilon_1+\varepsilon_2, \quad\quad\quad\quad\alpha_3=\varepsilon_2-\varepsilon_1, \\
& \alpha_4=\varepsilon_3-\varepsilon_2, \quad\quad\quad\quad \alpha_5=\varepsilon_4-\varepsilon_3, \\ 
& \alpha_6=\varepsilon_5-\varepsilon_4, \quad\quad\quad\quad \alpha_7=\varepsilon_6-\varepsilon_5 .
\end{aligned}
\end{equation}

\subsection{\texorpdfstring{$E_8$}{E8}}

The root basis vectors of $E_8$ type are
\begin{equation}\label{eq: e8roots}
\begin{aligned}
&\alpha_1=\frac{1}{2}\left(\varepsilon_1+\varepsilon_8\right)-\frac{1}{2}\left(\varepsilon_2+\varepsilon_3+\varepsilon_4+\varepsilon_5+\varepsilon_6+\varepsilon_7\right), \\
&\alpha_2=\varepsilon_1+\varepsilon_2, 
\quad\quad\quad\quad
\alpha_3=\varepsilon_2-\varepsilon_1,\\ &\alpha_4=\varepsilon_3-\varepsilon_2, 
\quad\quad\quad\quad
\alpha_5=\varepsilon_4-\varepsilon_3 \\
&\alpha_6=\varepsilon_5-\varepsilon_4, 
\quad\quad\quad\quad
\alpha_7=\varepsilon_6-\varepsilon_5,\\ &\alpha_8=\varepsilon_7-\varepsilon_6 .
\end{aligned}
\end{equation}

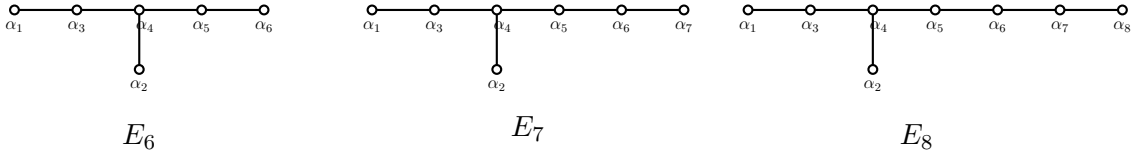
\begin{figure}[H]
\centering
\begin{minipage}{0.32\textwidth}
\centering
\begin{tikzpicture}[
  scale=0.55,
  transform shape,
  x=1.5cm, y=1.3cm,
  every node/.style={font=\large},
  vertex/.style={
    circle,
    draw,
    fill=white,
    inner sep=0pt,
    minimum size=6pt,
    line width=.8pt
  }
]
  
  \coordinate (a1) at (0,0);
  \coordinate (a3) at (1,0);
  \coordinate (a4) at (2,0);
  \coordinate (a5) at (3,0);
  \coordinate (a6) at (4,0);
  \coordinate (a2) at (2,-1.1);

  \draw[line width=.8pt] (a1) -- (a3) -- (a4) -- (a5) -- (a6);
  \draw[line width=.8pt] (a4) -- (a2);

  \foreach \p in {a1,a2,a3,a4,a5,a6}
    \node[vertex] at (\p) {};

  \foreach \p/\i in {a1/1,a3/3,a5/5,a6/6,a2/2}
    \node[below=4pt] at (\p) {$\alpha_{\i}$};

  \node[below=4pt, xshift=4pt] at (a4) {$\alpha_4$};
\end{tikzpicture}

\smallskip
\(E_6\)
\end{minipage}
\hfill
\begin{minipage}{0.32\textwidth}
\centering
\begin{tikzpicture}[
  scale=0.55,
  transform shape,
  x=1.5cm, y=1.3cm,
  every node/.style={font=\large},
  vertex/.style={
    circle,
    draw,
    fill=white,
    inner sep=0pt,
    minimum size=6pt,
    line width=.8pt
  }
]
  
  \coordinate (a1) at (0,0);
  \coordinate (a3) at (1,0);
  \coordinate (a4) at (2,0);
  \coordinate (a5) at (3,0);
  \coordinate (a6) at (4,0);
  \coordinate (a7) at (5,0);
  \coordinate (a2) at (2,-1.1);

  \draw[line width=.8pt] (a1) -- (a3) -- (a4) -- (a5) -- (a6) -- (a7);
  \draw[line width=.8pt] (a4) -- (a2);

  \foreach \p in {a1,a2,a3,a4,a5,a6,a7}
    \node[vertex] at (\p) {};

  \foreach \p/\i in {a1/1,a3/3,a5/5,a6/6,a7/7,a2/2}
    \node[below=4pt] at (\p) {$\alpha_{\i}$};

  \node[below=4pt, xshift=4pt] at (a4) {$\alpha_4$};
\end{tikzpicture}
\smallskip
\(E_7\)
\end{minipage}
\hfill
\begin{minipage}{0.32\textwidth}
\centering
\begin{tikzpicture}[
  scale=0.55,
  transform shape,
  x=1.5cm, y=1.3cm,
  every node/.style={font=\large},
  vertex/.style={
    circle,
    draw,
    fill=white,
    inner sep=0pt,
    minimum size=6pt,
    line width=.8pt
  }
]
  
  \coordinate (a1) at (0,0);
  \coordinate (a3) at (1,0);
  \coordinate (a4) at (2,0);
  \coordinate (a5) at (3,0);
  \coordinate (a6) at (4,0);
  \coordinate (a7) at (5,0);
  \coordinate (a8) at (6,0);
  \coordinate (a2) at (2,-1.1);

  \draw[line width=.8pt] (a1) -- (a3) -- (a4) -- (a5) -- (a6) -- (a7) -- (a8);
  \draw[line width=.8pt] (a4) -- (a2);

  \foreach \p in {a1,a2,a3,a4,a5,a6,a7,a8}
    \node[vertex] at (\p) {};

  \foreach \p/\i in {a1/1,a3/3,a5/5,a6/6,a7/7,a8/8,a2/2}
    \node[below=4pt] at (\p) {$\alpha_{\i}$};

  \node[below=4pt, xshift=4pt] at (a4) {$\alpha_4$};
\end{tikzpicture}

\smallskip
\(E_8\)
\end{minipage}

\caption{Dynkin diagrams of \(E_6\), \(E_7\), and \(E_8\).}
\label{fig:dynkin_E678}
\end{figure}
\section{Classification of shift vectors}\label{sec:classification}
In this appendix we classify inequivalent shift vectors $w$ for $E_8 \times E_8$ and $\Spin(32)/\Z_2$ while fixing $v=\frac12(0^{14},1,1)$. Let us first state what we mean by equivalent.

Modular invariance requires order 4 shifts $w \in \frac14 \Gamma_{0,16}$ altogether with $4w^2 \in 2\Z +1$ and $4w \cdot v \in 4\Z +1$. Two shifts $w,w'$ are equivalent if they differ by the following transformations:
\begin{align}
  w^\prime=s(w)+\alpha \, , \quad \alpha \in  \Gamma_{0,16} \, 
\end{align}
where $s$ is an element of Weyl subgroup of gauge algebra which stabilises $v$. 
We can check the equivalence of two shift vectors explicitly as follows. Let $A=cv+kw, B=dv+lw$, then the part of the partition function which comes from the lattice $\Gamma_{0,16}$ is
\begin{equation}
\bar\Gamma_{16,w}^{k,l}\chvec{c}{d}=e^{\pi \i A\cdot B}\sum_{p\in\Gamma_{0,16}}\bar q^{\frac12(p+A)^2}e^{-2\pi \i (p+A)\cdot B} \, .
\end{equation}
Replacing $w$ by $w^\prime=s(w)+\alpha$ yields
\begin{equation}
\begin{aligned}        \bar\Gamma_{16,w^\prime}^{k,l}=&e^{\pi \i(s(A)+k\alpha)\cdot (s(B)+l\alpha)}\sum_{p\in\Gamma_{0,16}}\bar q^{\frac12(p+s(A)+k\alpha)^2}e^{-2\pi i(p+s(A)+k\alpha)\cdot (s(B)+l\alpha)}\\
        =&e^{\pi \i(dk-lc)v\cdot\alpha}\bar\Gamma_{16,w}^{k,l} \, ,
\end{aligned}
\end{equation}
which leads to $\bar\Gamma_{16,w+\alpha}^{k,l} \equiv\bar\Gamma_{16,w}^{k,l}$ up to discrete torsion $e^{ \pi \i (dk-lc)\Z}$ \cite{Vafa:1994rv}. Here $v \cdot \alpha \in \Z$ follows from the mixed condition on the shift vectors. Here we used the fact that a Weyl group action preserves inner product: $s(A)\cdot s(B)=A\cdot B$, for example.

We will classify inequivalent shift vectors on $E_8 \times E_8$ using Kac's algorithm (see \cite{Font:2020rsk} and references therein). Given the structure of this algebra, the discussion can be carried out for each $E_8$ factor separately to combine the admissible pairs at the end. For $\Spin(32)/\Z_2$ it will be easier to use a different method instead. Recalling the definition of the $\Spin(32)/\Z_2$ lattice
\begin{equation}
     \Gamma_{\Spin(32)/\Z_2}=\left\{p\in\Z^{16}\text{ or }\qty(\Z+\tfrac{1}{2})^{16}\middle|\sum_{i=1}^{16}p_i \in2\Z\right\} \, ,
\end{equation}
we will take separately into account vectors of the form $4w \in \Z^{16}$ and $4w \in \left( \Z+\frac12 \right)^{16}$.

\subsection{\texorpdfstring{$E_8\times E_8$}{E8×E8}}
\subsubsection{On the first \texorpdfstring{$E_8$}{E8}}
We describe the \(E_8\times E_8\) shift classes by a convenient change of
simple-root basis.  The point is to enlarge the list of simple-root candidates
by adding the negative of the highest root.

Let \(\alpha_1,\ldots,\alpha_8\) be the simple roots of \(E_8\), with the
conventions used above.  We define \(\alpha_9\) by
\begin{equation}
  \alpha_9
  =
  -\left(
  2\alpha_1+3\alpha_2+4\alpha_3+6\alpha_4
  +5\alpha_5+4\alpha_6+3\alpha_7+2\alpha_8
  \right).
\end{equation}
Thus \(-\alpha_9\) is the highest root of \(E_8\).  Equivalently,
\(-\alpha_9\) is the positive root such that, for every positive root
\(\rho\), the difference
\begin{equation}
  -\alpha_9-\rho
\end{equation}
is a non-negative integral linear combination of
\(\alpha_1,\ldots,\alpha_8\).

Of course, \(\alpha_9,\alpha_1,\ldots,\alpha_8\) are not independent, but they
satisfy
\begin{equation}
  \alpha_9
  +2\alpha_1+3\alpha_2+4\alpha_3+6\alpha_4
  +5\alpha_5+4\alpha_6+3\alpha_7+2\alpha_8
  =0.
\end{equation}
We draw a graph whose nodes are the roots \(\alpha_i\), with an edge between
two nodes when the corresponding roots have inner product \(-1\).  In our
convention the graph is
\begin{equation}
\begin{array}{ccccccccccccccc}
\alpha_1 & - & \alpha_3 & - & \alpha_4 & - & \alpha_5 & - &
\alpha_6 & - & \alpha_7 & - & \alpha_8 & - & \alpha_9
\\[-1mm]
&&&& | \\
&&&& \alpha_2 .
\end{array}
\end{equation}
This is called the extended Dynkin diagram of $E_8$.

Let \(w\) be a \(\mathbb Z_4\) shift in one \(E_8\) factor.  Here
\(\alpha_i\cdot w\) means the ordinary Euclidean inner product of the root
\(\alpha_i\) with the shift vector \(w\).  Up to Weyl reflections and lattice
shifts, we choose \(w\) so that
\begin{equation}
  \alpha_i\cdot w\ge 0 \qquad (i=1,\ldots,8),
  \qquad
  -\alpha_9\cdot w\le 1.
\end{equation}
If $-\alpha_9\cdot w>1$, then an equivalent shift vector $w_n^\prime=w+n\alpha_9$ satisfies $-\alpha_9\cdot w^\prime=-\alpha_9\cdot w-2n<1$ for some integer $n$.
We then attach non-negative integers to the nodes by
\begin{equation}
  s_i=4\,\alpha_i\cdot w \qquad (i=1,\ldots,8),
  \qquad
  s_9=4(1+\alpha_9\cdot w).
\end{equation}
We record the shift class as
\begin{equation}
  [s_1s_2\cdots s_8;s_9].
\end{equation}
The relation among the roots implies
\begin{equation}
  2s_1+3s_2+4s_3+6s_4+5s_5+4s_6+3s_7+2s_8+s_9=4.
\end{equation}

The unbroken semisimple root system is obtained by keeping precisely the
nodes with label \(0\).  In other words, the zero-labelled part of the graph
is read as the Dynkin diagram of the unbroken semisimple algebra.

For the first \(E_8\) factor, ten shift classes are summarised in table \ref{tab: first comp E8}. 
\begin{table}[h]
    \centering
    \[
\begin{array}{c|c|c|c|c}
\text{class}
& [s_1\cdots s_8;s_9]
& \text{zero-labelled subgraph}
& 4\mathsf A_i^2 \bmod 2
\\
\hline
\mathsf A_1  & [00000000;4] 
     & E_8 & 0 \\
\mathsf A_2  & [00000002;0] 
     & E_7\oplus A_1 & 0 \\
\mathsf A_3  & [20000000;0] 
     & D_8 & 0 \\
\mathsf A_4  & [10000000;2]
     & D_7 & 1 \\
\mathsf A_5  & [00000100;0] 
     & D_5\oplus A_3 & 1 \\
\mathsf A_6  & [01000000;1] 
     & A_7 & 0 \\
\mathsf A_7  & [00000001;2] 
     & E_7 & 1/2 \\
\mathsf A_8  & [10000001;0] 
     & D_6\oplus A_1 & 1/2 \\
\mathsf A_9  & [00000010;1] 
     & E_6\oplus A_1 & 3/2 \\
\mathsf A_{10} & [00100000;0] 
     & A_7\oplus A_1 & 3/2
\end{array}
\]
    \caption{First eight components of $E_8\times E_8$ shift vectors.
    }
    \label{tab: first comp E8}
\end{table}

\subsubsection{On the second \texorpdfstring{$E_8$}{E8}}
For the second \(E_8\) factor, the fixed \(\mathbb Z_2\) shift \(v\)
first leaves an \(E_7\oplus A_1\) root system.  The relevant nonabelian part
is read from the \(E_7\) factor.  In this paragraph, we again denote the
simple roots of this \(E_7\) by
\(\alpha_1,\ldots,\alpha_7\).  We define
\begin{equation}
  \alpha_0
  =
  -\left(
  2\alpha_1+2\alpha_2+3\alpha_3+4\alpha_4
  +3\alpha_5+2\alpha_6+\alpha_7
  \right).
\end{equation}
Thus \(-\alpha_0\) is the highest root of this \(E_7\).  The roots
\(\alpha_0,\alpha_1,\ldots,\alpha_7\) satisfy
\begin{equation}
  \alpha_0
  +2\alpha_1+2\alpha_2+3\alpha_3+4\alpha_4
  +3\alpha_5+2\alpha_6+\alpha_7
  =0.
\end{equation}
Their graph is
\begin{equation}
\begin{array}{ccccccccccccc}
\alpha_0 & - & \alpha_1 & - & \alpha_3 & - & \alpha_4 & - &
\alpha_5 & - & \alpha_6 & - & \alpha_7
\\[-1mm]
&&&&&& | \\
&&&&&& \alpha_2 .
\end{array}
\end{equation}
This is called the extended Dynkin diagram of $E_7$.

For an order-four shift in this \(E_7\) factor, we choose \(w\) so that
\begin{equation}
  \alpha_i\cdot w\ge0 \qquad (i=1,\ldots,7),
  \qquad
  -\alpha_0\cdot w\le1,
\end{equation}
up to lattice shift, and define the node labels by
\begin{equation}
  r_i=4\,\alpha_i\cdot w \qquad (i=1,\ldots,7),
  \qquad
  r_0=4(1+\alpha_0\cdot w).
\end{equation}
We record the class as
\begin{equation}
  [r_0;r_1r_2\cdots r_7].
\end{equation}
The labels obey
\begin{equation}
  r_0+2r_1+2r_2+3r_3+4r_4+3r_5+2r_6+r_7=4.
\end{equation}
Again, the zero-labelled part of the graph gives the Dynkin diagram of the
unbroken semisimple algebra.

The ten \(\mathsf B_i\) classes are summarized in table \ref{tab: second comp E8}.
\begin{table}[H]
    \centering
    \[
\begin{array}{c|c|c|c|c}
\text{class}
& [r_0;r_1\cdots r_7]
& \text{zero-labelled subgraph}
& 4\mathsf B_i^2 \bmod 2
\\
\hline
\mathsf B_1  & [0;1000010] 
     & D_4\oplus A_1\oplus A_1 & 1 \\
\mathsf B_2  & [0;0000012] 
     & D_6 & 1 \\
\mathsf B_3  & [2;0000002] 
     & E_6 & 0 \\
\mathsf B_4  & [0;0200000] 
     & A_7 & 0 \\
\mathsf B_5  & [0;0000101]
     & A_5\oplus A_1 & 0 \\
\mathsf B_6  & [1;0100001] 
     & A_5 & 1/2 \\
\mathsf B_7  & [0;0000020] 
     & D_6\oplus A_1 & 1/2 \\
\mathsf B_8  & [4;0000000] 
     & E_7 & 1/2 \\
\mathsf B_9  & [2;0000010] 
     & D_5\oplus A_1 & 3/2 \\
\mathsf B_{10} & [0;0001000] 
     & A_3\oplus A_3\oplus A_1 & 3/2
\end{array}\]\caption{Second eight components of $E_8\times E_8$ shift vectors. }
 \label{tab: second comp E8}
\end{table}

Finally, writing the full \(E_8\times E_8\) shift as
\begin{equation}
  w=(\mathsf A;\mathsf B),
\end{equation}
the condition \(4w^2\in 2\mathbb Z+1\) is equivalent to
\begin{equation}\label{eq:here1}
  \mathsf A^2+\mathsf B^2\equiv 1 \pmod 2.
\end{equation}
Hence the allowed pairs are read off from the last columns of the two tables:
\begin{equation}
\begin{aligned}
&\{\mathsf A_1,\mathsf A_2,\mathsf A_3,\mathsf A_6\}
\times
\{\mathsf B_1,\mathsf B_2\},
\\
&\{\mathsf A_4,\mathsf A_5\}
\times
\{\mathsf B_3,\mathsf B_4,\mathsf B_5\},
\\
&\{\mathsf A_7,\mathsf A_8\}
\times
\{\mathsf B_6,\mathsf B_7,\mathsf B_8\},
\\
&\{\mathsf A_9,\mathsf A_{10}\}
\times
\{\mathsf B_9,\mathsf B_{10}\}.
\end{aligned}
\end{equation}
Then there are $24$ inequivalent \(E_8\times E_8\) shift classes.

\subsection{\texorpdfstring{$\Spin(32)/\Z_2$}{Spin(32)/Z2}}

\subsubsection{case 1}
In this subsection let us assume that $4w\in\Z^{16}$. In this case the components of $w$ in terms of standard basis on $\R^{16}$, which include $\Spin(32)/\Z_2$ lattice, would be 
\begin{equation}
    \pm\frac{1}{4}, \pm\frac{2}{4},\pm\frac{3}{4},\pm 1
\end{equation}
up to lattice shifts. Since $w$ should satisfy $4v\cdot w\in 4\Z+1$, the last two components of $w$ should be one of the following:
\begin{equation}
    \qty(\frac14,\frac14),\qty(0,\frac12)
\end{equation}
up to lattice shifts. For instance, one might take them to be $\qty(\frac{3}{4},-\frac14)$, but these can be shifted to $\qty(\frac14,\frac14)$ by $\frac12\qty(\cdots,-1,1)\in \Gamma_{\Spin(32)/\Z_2}$. 

\textit{Claim}. One can take the first $14$ components of $w$ as follows:
\begin{equation}
    \qty(0^{n_0},\qty(\pm\frac14)^{n_1},\qty(\pm \frac12)^{n_2};)
\end{equation}
up to lattice shifts. What differs from the $E_8\times E_8$ case is that if we keep this expression for a shift vector, the pair of numbers $(n_0,n_1,n_2)$ is invariant under lattice shifts and Weyl reflection.

Let us show this statement. We start from the form of the first $14$ components:
\begin{equation}
    \qty(0^{n_0},\qty(\pm\frac14)^{n_1},\qty(\pm \frac12)^{n_2},\qty(\pm \frac34)^{n_3}).
\end{equation}
Here we can assume that $n_3=0,1$, because if $n_3\geq 2$, there is at least one pair of two $\frac34$'s. Then they can be shifted to a pair of $\frac14^2$; e.g, $\qty(\cdots,\frac34^{2n})\to \qty(\cdots,\frac14^{2n})$ by $(\cdots,0,1^{2n})\in \Spin(32)/\Z_2$ lattice. 

Next, we assume that $n_3=1$:
\begin{equation}
    \qty(0^{n_0},\qty(\pm\frac14)^{n_1},\qty(\pm \frac12)^{n_2},\qty(\pm \frac34)^{1}).
\end{equation}
because if $n_3=0$, we obtain $\qty(0^{n_0},\qty(\pm \frac14)^{n_1},\qty(\pm\frac12)^{n_2})$ and the proof was already finished. Now we can assume that $n_0\geq 1$ or $n_2 \geq 1$, because otherwise $n_1=13$. Then the norm of this vector is 
\begin{equation}
    \qty(\qty(\pm\frac14)^{13},\pm \frac34)^2=\frac{11}{8},
\end{equation}
and can never yield $4w^2\in 2\Z+1$ with the norms of the last two components:
$\qty(\frac14,\frac14)^2=\frac18$ or $\qty(0,\frac12)^2=\frac14$. Now proof is finished and remained thing is to identify possible pair of $(n_1,n_2)$ with $n_0=14-n_1-n_2$ which satisfies $4w^2\in2\Z+1$. The result is 
\begin{equation}
    \begin{aligned}
    W_{0,\frac12}(n_1,n_2)=&
    \qty(\frac{1}{4}\qty(0^{14-n_1-n_2},1^{n_1},2^{n_2}),0,\frac{1}{2}):\\
    (n_1,n_2)\in\Bigl\{
(0,0),(0,2)&,(0,4),(0,6),
(4,1),(4,3),(4,5),
(8,0),(8,2),(12,1)
\Bigr\},\\
    \end{aligned}
\end{equation}
and
\begin{equation}
    \begin{aligned}
        W_{\frac14,\frac14}(n_1,n_2)=&\qty(\frac{1}{4}\qty(0^{14-n_1-n_2},1^{n_1},2^{n_2}),\frac14,\frac14):\\
(n_1,n_2)\in\Bigl\{
(2,0),(2,2)&,(2,4),(2,6),
(6,1),(6,3),(10,0),(10,2)
\Bigr\} \, .
    \end{aligned}
\end{equation}
and

\begin{equation}
    \widetilde W_{\frac14,\frac14}(n)=\qty(-\frac12,\frac12^{11-8n},\frac14^{2+8n},\frac14,\frac14),\quad n=0,1.\\
\end{equation}

\subsubsection{case 2}
In this subsection we take as $4w\in\qty(\Z+\frac12)^{16}$.
By a similar argument, we obtain the following four shift vectors \:
\begin{equation}
\begin{aligned}
    W_{\frac18,\frac38}(n)=&\left(\qty(\frac18)^{11-4n},\qty(\frac38)^{3+4n},\frac18,\frac38\right) ,\quad n=0,1,\\
   W_{\frac58,-\frac18}(n)=&
\left(
\left(\frac18\right)^{13-4n},
\left(\frac38\right)^{1+4n},
\frac58,
-\frac18
\right),
\quad n=0,1.\\
\end{aligned}
\end{equation}

Of course $W_{\frac18,\frac38}(2)$ and $W_{\frac58,-\frac18}(n), n=2,3$ are also allowed, but they are equivalent to some of the above four vectors up to Weyl reflection and lattice shifts:
\begin{equation}
    \begin{aligned}
        W^\prime =-W+\qty(\frac12^{16}).
    \end{aligned}
\end{equation}
Permutations of the components with even number of sign flip are allowed in the Weyl group of $D_n$.

\end{appendix}

\bibliographystyle{JHEP}
\bibliography{reference}

\end{document}